\documentclass[a4paper,11pt]{article}
\usepackage[utf8]{inputenc}

\usepackage[hidelinks]{hyperref}

\usepackage{framed}
\usepackage{mdframed}

\usepackage{setspace}

\usepackage{fullpage}
\usepackage[T1]{fontenc}

\usepackage{amsfonts}

\usepackage[british]{babel}
\usepackage{verbatim}
\usepackage[T1]{fontenc}
\usepackage{lmodern}
\usepackage{lipsum}
\usepackage{booktabs}
\usepackage{caption}
\usepackage{cite}
\usepackage{soul,color}
\usepackage[toc,page]{appendix}
\usepackage{tikz,pgfplots}

\usepackage{ifpdf}

\usepackage{amsthm, amssymb}
\usepackage[tbtags]{amsmath}
\usepackage{bm}
\usepackage{mathtools}
\usepackage{amstext}
\usepackage{braket}
\usepackage{multirow}
\usepackage{hyperref}

\usepackage[normalem]{ulem}
\usepackage{xcolor}

\begin{document}
\vspace{5mm}
\vspace{0.5cm}

\def\be{\begin{eqnarray}}
\def\ee{\end{eqnarray}}

\def\ba{\begin{aligned}}
\def\ea{\end{aligned}}

\def\rmd{\mathrm{d}}

\def\ls{\left[}
\def\rs{\right]}
\def\lc{\left\{}
\def\rc{\right\}}

\def\p{\partial}

\def\S{\Sigma}

\def\s{\sigma}

\def\O{\Omega}

\def\a{\alpha}
\def\b{\beta}
\def\g{\gamma}

\def\ad{{\dot \alpha}}
\def\bd{{\dot \beta}}
\def\gd{{\dot \gamma}}
\newcommand{\ft}[2]{{\textstyle\frac{#1}{#2}}}
\def\ib{{\overline \imath}}
\def\jb{{\overline \jmath}}
\def\Re{\mathop{\rm Re}\nolimits}
\def\Im{\mathop{\rm Im}\nolimits}
\def\trace{\mathop{\rm Tr}\nolimits}
\def\rmi{{ i}}
\def\N{\mathcal{N}}

\newcommand{\SU}{\mathop{\rm SU}}
\newcommand{\SO}{\mathop{\rm SO}}
\newcommand{\U}{\mathop{\rm {}U}}
\newcommand{\USp}{\mathop{\rm {}USp}}
\newcommand{\OSp}{\mathop{\rm {}OSp}}
\newcommand{\Symp}{\mathop{\rm {}Sp}}
\newcommand{\Sl}{\mathop{\rm {}S}\ell }
\newcommand{\Gl}{\mathop{\rm {}G}\ell }
\newcommand{\Spin}{\mathop{\rm {}Spin}}

\newcommand{\MS}[1]{\textcolor{red}{\textsf{[MS: #1]}}}
\newcommand{\JF}[1]{\textcolor{blue}{\textsf{[JF: #1]}}}
\newcommand{\GN}[1]{\textcolor{green}{\textsf{[GN: #1]}}}

\def\hc{c.c.}

\numberwithin{equation}{section}

\allowdisplaybreaks

\allowbreak

\setcounter{tocdepth}{2}


\begin{titlepage}
	\thispagestyle{empty}
	\begin{flushright}

		\hfill{LMU-ASC 22/23}
		
		\hfill{MPP-2023-135}

	\end{flushright}
\vspace{35pt}

	\begin{center}
	      
	     { \LARGE\bf{
	Cosmic Acceleration and Turns in the Swampland
	     }}

		\vspace{50pt}

		{\large Julian Freigang$^{1}$, Dieter~L\"ust$^{1,2}$, Guo-En Nian$^{3}$ and Marco~Scalisi$^{1}$}

		\vspace{25pt}

		{
			$^1${\it Max-Planck-Institut f\"ur Physik (Werner-Heisenberg-Institut),\\ F\"ohringer Ring 6, 80805, M\"unchen, Germany }

		\vspace{15pt}
            $^2${\it Arnold-Sommerfeld-Center for Theoretical Physics,\\ Ludwig-Maximilians-Universit\"at, 80333 M\"unchen, Germany }

\vspace{15pt}
            $^3${\it Institute for Theoretical Physics,\\ Utrecht University, Princetonplein 5, 3584 CE Utrecht, The Netherlands}
 		}
		\vspace{40pt}

		{ABSTRACT}
	\end{center}

	\vspace{10pt}
\noindent
We argue that field trajectories, which lead to cosmic acceleration and feature rapid turns near the boundary of the moduli space, are in the Swampland. We obtain this result by assuming the validity of the Swampland Distance Conjecture (SDC) in the presence of a positive scalar potential and by focusing on hyperbolic spaces, as prototype geometries of infinite distance limits of Calabi–Yau compactifications. We find that, in a quasi-de Sitter space with Hubble rate $H$ and acceleration parameter $\epsilon$, the turning rate $\Omega$ is upper bounded such as $\Omega/H<\mathcal{O}(\sqrt{\epsilon})$. Therefore, field trajectories consistent with the SDC can only have a negligible deviation from geodesics. This has direct implications for the realization and consistency of multi-field scenarios in string theory. Moreover, it implies a tension between asymptotic accelerating expansion, consistent with observations, and the de Sitter conjecture.

\bigskip

\end{titlepage}


\def\thefootnote{\fnsymbol{footnote}}

\vskip 0.5cm

\vspace{0.5cm}

\def\thefootnote{\arabic{footnote}}
\setcounter{footnote}{0}

\baselineskip 5.6 mm



\tableofcontents


\newpage


\section{Introduction}
\label{sec:Intro}
Cosmic acceleration plays a fundamental role in the current understanding of our Universe. A variety of experiments operating at different scales, such as supernovae \cite{SupernovaSearchTeam:1998fmf,Riess:2021jrx} and cosmic microwaves background (CMB) \cite{Boomerang:2000jdg, SDSS:2003eyi, Aghanim:2018eyx} experiments, have provided very compelling evidence of a phase of accelerating expansion both in the early and in current universe. While we have a good understanding of how this phase can be realized in terms of an effective scalar field theory, we still struggle to agree on a full-fledged embedding of it in string theory.

The presence of several light scalar fields, active during the acceleration phase,\footnote{In this work, we consider only time-dependent acceleration phase, such as inflation or quintessence, with certain displacements in field space.} is a natural expectation for such an embedding (see \cite{Cicoli:2023opf} for a recent review on string cosmology). String theory comes in fact with many moduli, often spanning non-trivial field geometries, and giving them a mass is definitely a complex task. Unlike scenarios with one single scalar field, multi-field models typically feature {\it non-geodesic trajectories} in field space.\footnote{It should be noted that the most common strategy to construct an effective (supergravity) model is to stabilize all fields except one, which drives the acceleration phase. However, despite its simplicity, this approach may not be the most natural and often demands precise control over the effective theory.} Deviations from geodesics can be sourced by a non-zero scalar potential and they are usually quantified by the so-called {\it turning rate} $\Omega$. Strong non-geodesic motion, characterized by rapid turns in field space with $\Omega\gg 1$ (in Hubble units), can lead to intriguing and rich phenomenology. Examples have been provided in the context of inflation (see, e.g., \cite{Chakraborty:2019dfh,Dimopoulos:2005ac,Wands:2007bd,Cremonini:2010ua,Yang:2012bs,Brown:2017osf,Christodoulidis:2018qdw,Dias:2018pgj,Achucarro:2018vey,Aragam:2019omo,Aragam:2020uqi,Aragam:2021scu,Renaux-Petel:2021yxh,Bhattacharya:2022fze}) and also for quintessence models (see, e.g., \cite{Cicoli:2020cfj,Cicoli:2020noz,Akrami:2020zfz,Anguelova:2021jxu,Eskilt:2022zky,Brinkmann:2022oxy,Shiu:2023nph,Shiu:2023rxt}). These can be modifications of the inflationary power spectrum \cite{Cremonini:2010ua,Brown:2017osf}, production of primordial black holes \cite{Palma:2020ejf,Fumagalli:2020adf,Anguelova:2020nzl}, possibility to inflate on a steep potential \cite{Achucarro:2018vey,Akrami:2020zfz,Eskilt:2022zky} (namely with large potential gradient) and also enhanced growth of large-scale structure \cite{Akrami:2020zfz}. Despite the great attention this topic gained in the research community, the results are mainly model-dependent and so we lack a general principle of what a consistent quantum gravity embedding allows for (see \cite{Aragam:2021scu} for some work in this direction in the context of supergravity). 

An alternative route to (string) model building is given by the Swampland program \cite{Vafa:2005ui,Ooguri:2006in,Palti:2019pca,vanBeest:2021lhn}, which suggests that one can employ a bottom-up approach to restrict the set of effective field theories (EFTs) consistent with quantum gravity. This is based on a number of universal consistency constraints, which act already at energies typically lower than the Planck mass $M_{\rm P}$, thus making them  meaningful for phenomenology.
One property that consistent EFTs appear to possess is a finite range of validity in field space. This is indeed one of the powerful implications of the {\it Swampland Distance Conjecture} (SDC) \cite{Ooguri:2006in}, which states that infinite scalar field variations are always accompanied by (at least) one infinite tower of states with exponentially decreasing mass scale. In this limit, the quantum gravity cut-off, which we identify with the species scale $\Lambda_{\text{s}}$ \cite{Dvali:2007hz,Dvali:2007wp,Dvali:2009ks,Dvali:2010vm,Dvali:2012uq}, decays exponentially in field space, thus leading to a breakdown of the effective theory.\footnote{It has also been pointed out that the limits of small (AdS) cosmological constant \cite{Lust:2019zwm}, small gravitino mass \cite{Cribiori:2021gbf,Castellano:2021yye,Anchordoqui:2023oqm} and small/large entropy \cite{Bonnefoy:2019nzv,Cribiori:2022cho,Delgado:2022dkz,Cribiori:2023ffn} lead to analogous conclusions.} Field displacements of order $M_{\rm P}$ are enough to observe this behaviour \cite{Baume:2016psm, Klaewer:2016kiy} and to extract consequences relevant for phenomenology (see also \cite{Rudelius:2023mjy}). Implications of the SDC for cosmic inflation were first studied in \cite{Scalisi:2018eaz} (see also \cite{Scalisi:2019gfv,Bravo:2020wdr}), where a universal upper bound on the inflaton range was found (see also \cite{Etheredge:2022opl,vandeHeisteeg:2023uxj} for some variations of it with fixed decay rate).  

The SDC finds a natural test around the {\it boundary of moduli space}. These asymptotic regions are located at an infinite distance from any other point, hence referred to as `infinite-distance singularities'. Around these regions, the geometry exhibits negative curvature and non-compactness\footnote{In the context of inflationary cosmology, it has been shown \cite{Roest:2013aoa,Kallosh:2013yoa,Burgess:2014tja,Burgess:2014oma,Roest:2015qya,Burgess:2020qsc} that non-compact symmetries and negative curvature of the field space are key features for an excellent fit to the observational data. The relation between the SDC and such a cosmological scenario has  in fact been investigated in \cite{Scalisi:2018eaz}.}, while maintaining a finite volume \cite{Ooguri:2006in}. The effective theory becomes simple and can be expressed as a perturbative expansion on a certain parameter. Additionally, there is evidence suggesting that the scalar potential approaches zero in this limit \cite{Ooguri:2018wrx,Hebecker:2018vxz}. These factors have led to serious consideration of the boundary of moduli space as a promising framework for embedding models of cosmic acceleration \cite{Rudelius:2021azq,Rudelius:2022gbz,Calderon-Infante:2022nxb,Marconnet:2022fmx,Apers:2022cyl,Shiu:2023nph,Shiu:2023rxt,Cremonini:2023suw}, often referred to as `asymptotic acceleration'.

In this work, we study the implications of the SDC for multi-field models of cosmic acceleration at the boundary of moduli space. As main result, we find that the ratio between the turning rate $\Omega$ and the Hubble parameter $H$ is constrained by
\begin{equation}\label{OmegaBound}
    \frac{\Omega}{H}<c \ \sqrt{\epsilon}\,,
\end{equation}
with $\epsilon\equiv -\dot{H}/H^2$ being the acceleration parameter and $c$ being a $\mathcal{O}(1)$ quantity, depending on the curvature of the moduli space and on the decay rate of the tower mass scale. Since $\epsilon<1$ by definition, this result implies that asymptotic acceleration is incompatible with rapid turns or any strong non-geodesic motion. At the boundary of moduli space, quantum gravity imposes predominantly geodesic motion. We argue that this result should be valid also in the more conservative case of super-Planckian excursions, for which one can consistently apply the SDC. 

One direct implication of eq.~\eqref{OmegaBound} is a clear tension between asymptotic acceleration and the de Sitter conjecture \cite{Obied:2018sgi,Garg:2018reu,Ooguri:2018wrx}. In fact, a distinctive characteristic of multi-field models is that the acceleration phase is not solely determined by the gradient of the scalar potential, but rather by the interplay of this and the turning rate, as given by the following formula\footnote{Let us remark that eq.~\eqref{MFepsilon} relies on a slow-roll approximation, which assumes that the second derivative of the fields is sub-dominant compared to the friction term in the equations of motion. The full formula, as discussed in sec.~\ref{SecMF}, reveals that relaxing this condition can potentially aid in satisfying the de Sitter conjecture in an accelerating background.}:
\begin{equation}\label{MFepsilon}
    \epsilon=\frac12\frac{|\nabla V|^2}{ V^2}\left(1+\frac{\Omega^2}{9 H^2}\right)^{-1}\,.
\end{equation}
It has been previously highlighted in the literature \cite{Achucarro:2018vey} that, when $\Omega\gg H$, this formula allows for the fulfillment of the de Sitter conjecture (i.e., $|\nabla V| > \mathcal{O}(1) V$) while also enabling an acceleration phase with $\epsilon\ll1$. However, our result eq.~\eqref{OmegaBound} significantly limits this possibility within the context of asymptotic acceleration, since it implies that the second term in the bracket of eq.~\eqref{MFepsilon} is sub-leading. Given the current observational bounds, we conclude that models of early/late-time acceleration, near the boundary of moduli space, typically exhibit tension with the de Sitter conjecture.

The paper is organized as follows. In sec.~\ref{sec:SDCmoduli}, we introduce the SDC and show how the tower mass decay rate can strictly depend on the deviations from geodesic trajectories in field space. In sec.~\ref{sec:Trajctories}, we provide a pedagogical discussion of the multi-field framework and introduce the turning rate. In sec.~\ref{sec::Omega}, we investigate the case of infinite-distance trajectories with constant geodesic deviation and obtain our main result eq.~\eqref{OmegaBound}. We focus on hyperbolic field geometries, as prototype manifolds of the moduli space boundary for Calabi–Yau compactifications. In sec.~\ref{sec::nonconstOmega}, we extend our result to the case of infinite-distance trajectories with a time-dependent deviation angle from geodesics. In sec.~\ref{sec:concl}, we provide our conclusions. Throughout the text, we work in reduced Planck mass units ($M_{P}=1$).

\section{SDC, mass decay rate and non-geodesics}
\label{sec:SDCmoduli}

The Swampland Distance Conjecture (SDC) \cite{Ooguri:2006in} implies the existence of at least one infinite tower of states with mass scale exponentially decreasing in field space, in the infinite distance limit, namely
\begin{equation}\label{mSDC}
m=m_0 \exp(-\lambda \Delta )\qquad \text{as} \qquad \Delta\rightarrow\infty\,,
\end{equation}
where $m_0$ is the typical mass scale of the tower before any displacement, $\Delta$ is the traversed distance in moduli space and $\lambda$ is the \textit{decay rate}, namely the parameter regulating how fast the mass of the tower decreases in field space. It has been argued that $\lambda$ is order one \cite{Klaewer:2016kiy}, in reduced Planck mass units, and lower bounds have also been pointed out in different contexts \cite{Andriot:2020lea,Gendler:2020dfp,Lanza:2021udy,Castellano:2021yye,Etheredge:2022opl}. The existence of a lower bound is very important as it  defines the validity of the EFT. Namely, it indicates how fast/slow one can approach the infinite distance singularity in field space, and therefore how much field distance can be traversed, before the EFT completely breaks down, due to genuine quantum gravity effects. It happens, in fact, that the quantum gravity cut-off $\Lambda_{\text{s}}$, namely the species scale \cite{Dvali:2007hz,Dvali:2007wp,Dvali:2009ks,Dvali:2010vm,Dvali:2012uq}, decreases exponentially in field space, together with the mass scale of the tower. While a finite small number of light states can always be integrated in, such to define a new EFT, the presence of an {\it infinite} number of light states necessarily drives the cut-off to zero with exponential rate $\gamma$, which is in general different from the rate $\lambda$ of the tower. In the case of states equally spaced, such as Kaluza-Klein modes, one can show that  $\Lambda_{\text{s}}=m^{1/3}$ (assuming $M_{\text{P}}=1$), thus yielding $\gamma=\lambda/3$ \cite{Grimm:2018ohb,Hebecker:2018vxz,Scalisi:2018eaz,Cribiori:2021gbf}. This is consistent with the fact that, in the infinite distance limit, the quantum gravity cut-off lies still above the typical mass scale of the tower. While traversing a distance in field space, some states of the tower can enter the EFT and produce observational effects, while the quantum gravity cut-off remains still above the typical energy scale of the EFT.

The exponential rate of the tower $\lambda$ can in general depend on the {\it path} followed in moduli space to approach the infinite-distance point. A first example of this situation was given in \cite{Scalisi:2018eaz} for the hyperbolic half-plane, where it was shown that trajectories, with the axion and saxion linear to each other\footnote{Situations where the axion has a typical linear backreaction with the saxion, for large field displacements, have been observed in string theory models such as in \cite{Blumenhagen:2015qda,Baume:2016psm,Valenzuela:2016yny,Blumenhagen:2018hsh,Grimm:2020ouv}.},  yield an effective reduction of the decay rate.  This translates also into the possibility of engineering a larger field excursion. A more general analysis is given in \cite{Calderon-Infante:2020dhm}. In fact, one can reverse eq.~\eqref{mSDC} and express the {\it mass decay rate} of the tower as
\begin{equation}
\lambda(\Delta)= - \frac{\rmd \log m}{\rmd \Delta}=- T^i \partial_i \log m\,, \label{eq::generaldecayrate}
\end{equation}
that is the scalar product between the normalized tangent vector $T^i$, along the trajectory that we follow to reach the point at infinity, and the gradient of the (logarithm of the) mass of the tower. In the most general case, the gradient of the mass can be aligned along any direction in moduli space \cite{Calderon-Infante:2020dhm}. However, in most of the string theory examples, $\partial_i \log m$ is aligned along geodesics. This implies that $\lambda$ becomes a measure to quantify the {\it non-geodicity} of the trajectory. In this case, we can write 
\begin{equation}\label{exprate}
\lambda= - |\partial \log m| \cos\theta= \lambda_{\text g} \cos\theta\,,
\end{equation}
where $\theta$ is the angle between the trajectory we are following in field space and the geodesic.  Both paths will reach the infinite-distance singularity but with different angles. The parameter $\lambda_{\text g}$ represents the highest value of $\lambda$ and it corresponds in fact to the decay rate for a geodesic trajectory. Moving along a non-geodesic trajectory can be the result of introducing a scalar potential for the moduli (see Sec.~\ref{sec:Trajctories}).

\begin{figure}[t]
	\begin{center}
		\includegraphics[scale=0.58]{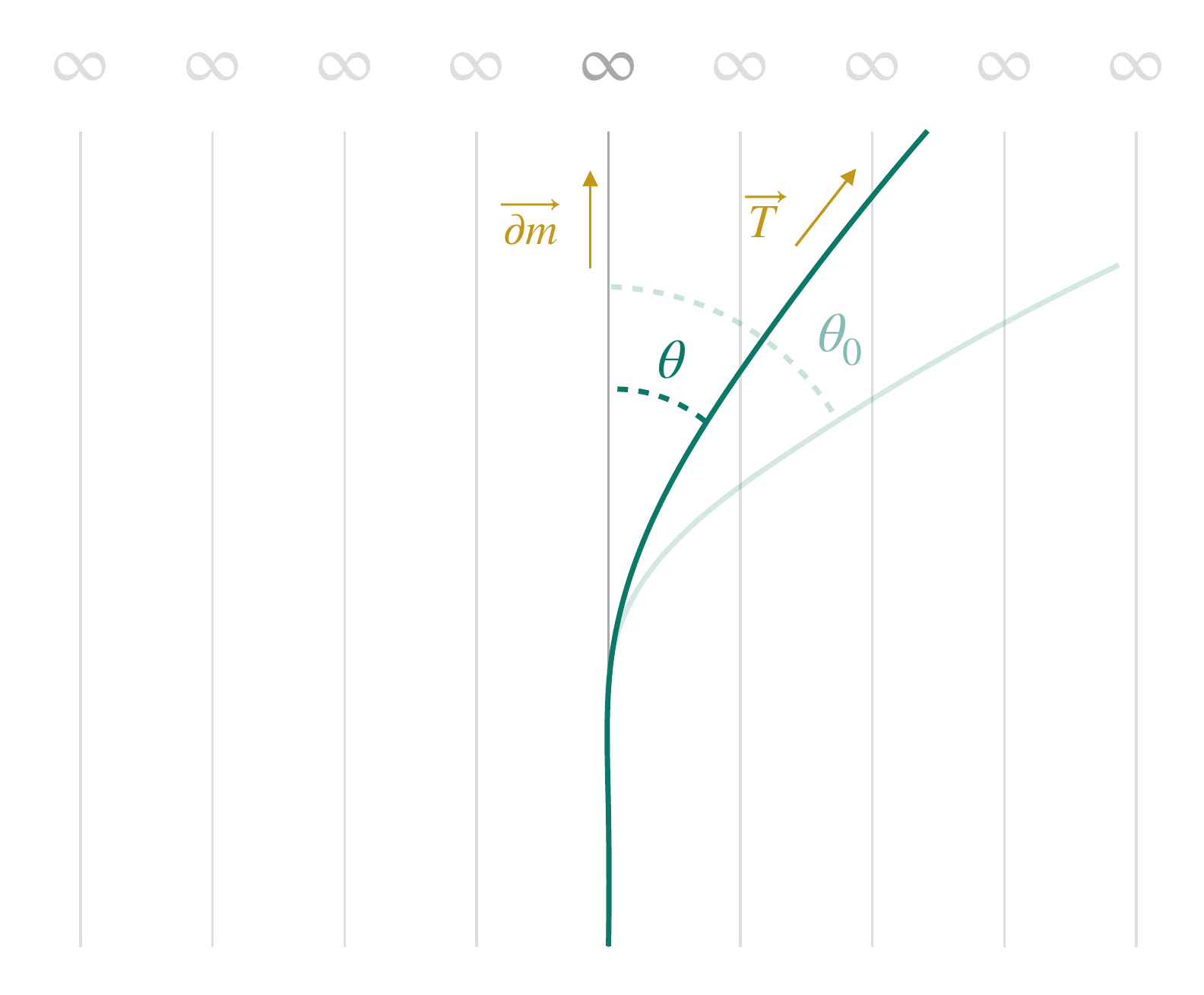}
		\vspace*{-0.2cm}\caption{\it Cartoon picture of a non-geodesic trajectory with tangent vector $\vec{T}$ deviating from the geodesic by an angle $\theta$. The set of infinite-distance geodesics is represented as parallel gray lines. The gradient of the tower mass $\vec{\partial m}$ aligns with the geodesic. Both the geodesic and the non-geodesic trajectories  approach the infinite distance region. The angle $\theta_0$ represents the maximum deviation, which is constrained by the lower bound on the decay rate of the SDC tower mass.}
		\label{cartoonSDC}
	\end{center}
\end{figure}

Eq.~\eqref{exprate} seems to suggest that $\lambda$ could even become zero, if one moves along a trajectory, which is orthogonal to a geodesic (i.e. $\theta=\pi/2$). This would mean that arbitrary distances could be traversed without the mass scale of a tower dropping-off. The EFT would be valid for arbitrary long distances, and one could easily avoid the drastic implication of the SDC.\footnote{Models with highly curved trajectories and large field ranges have in fact been proposed in literature \cite{Brown:2017osf,Achucarro:2018vey,Bjorkmo:2019aev,Aragam:2019omo,Cicoli:2020noz,Akrami:2020zfz,Aragam:2020uqi}. Whether these effective scenarios could be realized in a consistent string theory embedding is still unclear. Recent work \cite{Aragam:2021scu} seems in fact to restrict such a possibility.} However, as mentioned above, we have clear indications that string theory sets a lowest possible value for such a decay rate \cite{Andriot:2020lea,Gendler:2020dfp,Lanza:2021udy,Castellano:2021yye,Etheredge:2022opl}. If we generically indicate the existence of such a lower bound with
\begin{equation}\label{2.4}
\lambda\geq\lambda_0\,,
\end{equation}
then we can translate this into a maximum deviation angle $\theta_0$ from the geodesic trajectory allowed by the SDC, namely
\begin{equation}
\cos\theta\geq\cos\theta_0=-\frac{\lambda_0}{|\partial \log m|} =\frac{\lambda_0}{\lambda_{\text g} }\,.\label{eq::deviationangle}
\end{equation}
A bound on the angle $\theta$ means that not all the trajectories in moduli space are allowed by the SDC and can deviate by a maximum angle from the geodesic (see fig.~\ref{cartoonSDC}). In the next section, we recall how the introduction of a scalar potential can lead to a departure from a moduli space geodesic equation.

At this juncture, it is important to emphasize that our focus will now be solely on \textit{infinite-distance trajectories} in the subsequent discussion. These trajectories are characterized by distances that can extend infinitely, providing a robust framework to apply the SDC. 
Specifically, we will first examine deviations from geodesics with a constant angle\footnote{The situation of trajectories with constant deviation angle from a geodesic has been named `critical case' in \cite{Calderon-Infante:2020dhm}. In a 2-dimensional hyperbolic space, it corresponds to a linear backreaction between the saxion and the axion, as discussed in \cite{Scalisi:2018eaz} and in sec.~\ref{sec:oneHyp}.} $\theta=\text{const}$ (sec.~\ref{sec::Omega}) and then trajectories with a time-dependent deviation angle $\theta=\theta(t)$ (sec.~\ref{sec::nonconstOmega}). Let us emphasize that, given eq.~\eqref{exprate}, the latter case corresponds to a time-dependent, or rather $\Delta$-dependent, decay rate of the tower mass $\lambda=\lambda(\Delta)$. This will effectively induce field-dependent corrections such that the mass formula eq.~\eqref{mSDC} will deviate from its the exponential form when moving away from the moduli space boundary (which is placed at $\Delta\rightarrow\infty$). Therefore, a time-dependent decay rate can serve as a convenient means to parameterize a departure from the boundary.

\section{Multi-scalar field setup and trajectories in moduli space}
\label{sec:Trajctories}

String theory comes naturally with many moduli, namely massless scalar fields. The internal field geometry, defined by their kinetic terms, is generically non-flat, as the result of the compactification process, and characterized by a set of geodesics. However, the introduction of a scalar potential (e.g. by means of fluxes) can lead to a deviation from the original geodesic trajectories and a consequent change of dynamics.\footnote{There are instead situations where the dynamics remains quite insensitive to a great variety of scalar potentials and is instead mainly determined by the geometric properties of the internal manifold. In the context of inflationary cosmology, the $\alpha$-attractor scenario  \cite{Roest:2013aoa,Kallosh:2013yoa,Roest:2015qya,Scalisi:2015qga} is a primary example of such a circumstance.}

In this section, we present a pedagogical introduction to a convenient framework for studying multi-scalar field systems \cite{Achucarro:2018vey}. This is based on projecting the equations of motion along the tangent and normal directions of the trajectory along which the system evolves. By employing this approach, we demonstrate how the system can be described using an equation resembling the equation of motion for a single scalar field, as well as another equation involving the turning rate $\Omega$. We emphasize the relationship between the equations of motion and the trajectories (geodesic or non-geodesic) in field space. The presentation includes increasing levels of complexity. We begin by considering the case of free massless scalar fields in a flat Minkowski background in sec.~~\ref{sec:SFMink}. Next, we introduce a scalar potential and demonstrate how it leads to deviations from geodesic motion in sec.~\ref{sec:SFMinkV}. Subsequently, we incorporate gravity and investigate the effects of a Friedmann-Lema\^{i}tre-Robertson-Walker (FLRW) background in sec.~\ref{sec:SFFRWV}. Finally, in sec.~\ref{SecMF}, we describe the equations governing a multi-field system that gives rise to cosmic acceleration.

\subsection{Scalar fields in Minkowski spacetime}\label{sec:SFMink}

Let us consider the  Lagrangian of $n$ free massless homogeneous scalar fields  $\Phi^a=\Phi^a(t)$, thus depending just on the time variable $t$:

\begin{equation}
    \mathcal{L} = -\frac{1}{2} \eta^{\mu\nu} G_{ab}\ \partial_\mu \Phi^a \partial_\nu \Phi^b\,,\label{freeL}
\end{equation}
where $\eta^{\mu\nu}$ is the Minkowski spacetime metric, $G_{ab}=G_{ab}(\bm{\Phi})$ is the internal field space metric, and Latin indices $a,b$ run from 1 to $n$.
The equations of motion then take the form 
\begin{equation}
    \ddot{\Phi}^a+\Gamma^a_{bc} \dot{\Phi}^b \dot{\Phi}^c=0 \,,\label{eq::geodeq}
\end{equation}
where the dot $~\dot{}~$ is indicating a time derivative, and
\be
\Gamma^{a}_{bc} = \frac{1}{2} G^{ad}\left(\frac{\partial G_{bd}}{\partial \Phi^c}+\frac{\partial G_{cd}}{\partial \Phi^b}- \frac{\partial G_{bc}}{\partial \Phi^d}\right)
\ee
are the Christoffel symbols of the moduli space. Note that eq.~\eqref{eq::geodeq} has precisely the form of a {\it geodesic equation}. It describes in fact the set of (geodesic) trajectories along which the scalar fields $\Phi^a$ evolve in time. Notice that the time $t$ is not a preferred parameter for the geodesic and we can shift and rescale it as the result of the shift-symmetry of the Lagrangian eq.~\eqref{freeL} and scale-symmetry of the equations of motion eq.~\eqref{eq::geodeq}.

Let us introduce the covariant derivative $D_t$, which is defined as
\begin{equation}
    D_t A^a \equiv \dot{A}^a + \Gamma_{bc}^a A^b \dot{\Phi}^c\,,
\end{equation}
for a given vector $A^a$. Having $D_t A^a=0$ means that the vector $A^a$ is parallel transported along the trajectory $\Phi^a$, i.e. $A^a$ always `points at the same direction' along $\Phi^a$. Moreover, $D_t$ acting on a field scalar reduces to an ordinary time derivative. With this definition, the above set of equations \eqref{eq::geodeq} reduces to
\begin{equation}
    D_t \dot{\Phi}^a =0\,,\label{eq::geodeqCov}
\end{equation}
which is consistent with the fact that the equations of motion are just geodesic equations and a geodesic is a trajectory which is autoparallel transported along itself.

We now introduce  the tangent and the normal vector to the trajectory $\Phi^a$, respectively, as
\begin{eqnarray}
	T^a &=& \frac{\dot{\Phi}^a}{\dot{\Phi}}\\
	N^a &=& - \frac{1}{|D_t T|} D_t T^a \,,
\end{eqnarray}
where $\dot{\Phi}$ is the speed along the trajectory, defined as
\begin{equation}
    \dot{\Phi} = \sqrt{G_{ab}\dot{\Phi}^a \dot{\Phi}^b}\,.
\end{equation}
Both vectors $T^a$ and $N^a$ are normalised and orthogonal to each other, namely $G_{ab}T^aT^b =G_{ab}N^aN^b = 1$  and $G_{ab}T^aN^b = 0$. Now we can project the equations of motion along the tangent and normal vectors. This just means contracting the equations of motion with $T^a$ and $N^a$. The tangential projection yields
\begin{equation}
    \ddot{\Phi} = 0\,,
\end{equation}
where we have used the product rule for $D_t$ and the orthogonality property of $T^a$ and $N^a$. Instead, the normal projection gives
\begin{equation}
    \Omega \dot{\Phi} = 0\,,\label{normalEQ}
\end{equation}
where we have introduced the {\it turning rate} as 
\begin{equation}
\Omega = |D_t T|\,,
\end{equation}
and used again orthogonality of the two vectors together with the fact that
\begin{equation}
    D_t \dot{\Phi}^a = \dot{\Phi} D_t T^a + \ddot{\Phi} T^a\,. \label{eq:Relation}
\end{equation}
Excluding the trivial case $\dot{\Phi} = 0$, eq.~\eqref{normalEQ} implies that we need $\Omega = 0$ in order to fulfill the geodesic equation. That is the reason why $\Omega$ is also called non-geodesity factor. Furthermore, let us note that if $T^a$ gets parallel transported along $\Phi^a$, i.e. $D_t T^a = 0$, we immediately get $\Omega = 0$ and the equation of motion just reduces to the equation  $\ddot{\Phi} =0$. Since we are dealing with positive definite Riemannian field manifolds, the statement $\Omega = 0$ is equivalent to $D_tT^a=0$. Geometrically, $\Omega$ measures the failure of $T^a$ being parallel transported along $\Phi^a$.\\

\subsection{Scalar fields with potential in Minkowski spacetime}\label{sec:SFMinkV}

As next step, we now introduce a potential $V(\Phi^a)$ for the scalar fields. We still consider still flat Minkowski background such that the Lagrangian $\mathcal{L}$ becomes
\begin{equation}
    \mathcal{L} = -\frac{1}{2} \eta^{\mu\nu} G_{ab} \partial_\mu \Phi^a \partial_\nu \Phi^b - V(\Phi^a)\,.
\end{equation}
The equations of motion hence read
\begin{equation}
    D_t \dot{\Phi}^a +G^{ab} V_b =0\,, \label{eq::eomPot}
\end{equation}
where we define $V_b \equiv \partial V/\partial \Phi^b$. Projecting again the set of equations in the tangential and normal direction we get
\begin{eqnarray}
    \ddot{\Phi} +V_T &=& 0\,, \label{eq::eomPotTan}\\ 
    \Omega \dot{\Phi}  &=& V_N\,,  \label{eq::eomPotNorm}
\end{eqnarray}
where we have introduced $V_T \equiv T^a V_a$ and $V_N \equiv N^aV_a$, i.e., the corresponding projections of the gradient of the potential $V$. It is interesting to understand what happens in the case $D_t T^a = 0$, which is, as explained earlier, equivalent to $\Omega =0$. Rearranging the relation \eqref{eq:Relation}, we get
\begin{equation}
    D_t T^a = \frac{1}{\dot{\Phi}} \left( D_t \dot{\Phi}^a - T^a \ddot{\Phi}   \right)\,.\label{eq::geodesicdeviation}
\end{equation}
In the case of zero potential, as seen before, both terms in the bracket of the last equation vanish, thus automatically leading to turning rate $\Omega=0$. Instead, in the presence of a non-zero potential, the situation is slightly more involved. The two terms can in fact cancel each other, so that the trajectory will follow a geodesic path in field space. However, the acceleration along the trajectory will be still determined by the tangential projection of $V$, as expressed in eq.~\eqref{eq::eomPotTan}. We can get more insight about this situation by using the equations of motion \eqref{eq::eomPot} and \eqref{eq::eomPotTan} and rewriting eq.~\eqref{eq::geodesicdeviation} as
\begin{equation}
    D_t T^a = -\frac{1}{\dot{\Phi}} \left( G^{ab}V_b - T^a V_T  \right)\,.
\end{equation}
In order to have $\Omega = 0$, we immediately see that the gradient of the potential $V_a$ and the tangent vector $T_a$ have to be aligned. This means that geodesic trajectories are always characterized by a zero normal component of the scalar potential, namely $V_N=0$. Intuitively, the trajectory corresponds to a valley of the scalar potential. If there is no normal force, the tangent vector gets parallel transported along the trajectory. This is an approach used very often in string/supergravity model building as it hugely simplifies the analysis of the system. In a typical axion-saxion system, it corresponds to stabilize one of the two fields and leave the other very light to drive the acceleration phase. On the other hand, the multi-field framework allows, in principle, also for very sharp turns, $\Omega\gg1$, which means a great misalignment between the potential gradient flow and the tangent vector $T^a$.

\subsection{Scalar fields with potential in FLRW spacetime}\label{sec:SFFRWV}

We further generalise our setup by taking the 4d spacetime to be a FLRW metric $g_{\mu\nu}$ with line element of the form
\begin{equation}
    \rmd s^2 = -\rmd t^2 + a^2(t) \rmd\bm{x}^2\,.
\end{equation}
Then, given the action
\begin{equation}
    S = \int \rmd^4 x \sqrt{-g}\left( \frac{1}{2}R-\frac{1}{2} g^{\mu\nu} G_{ab} \partial_\mu \Phi^a \partial_\nu \Phi^b - V(\Phi^a)\right)\,,
\end{equation}\label{eq::actionFRLW}
with $g$ being the determinant of $g_{\mu\nu}$ and $R$ the Ricci scalar, we get the following equations
\begin{equation}
    D_t \dot{\Phi}^a + 3H \dot{\Phi}^a + G^{ab} V_b = 0 \,.\label{eq::eomInf}
\end{equation}
These contain an addition friction term, proportional to the Hubble expansion rate $H \equiv\dot{a}/{a}$. The projections work completely analogous to the previous cases, namely we get
\begin{eqnarray}
    \ddot{\Phi} + 3H \dot{\Phi}+ V_T &=& 0\,, \label{eq::eomPHIdotV} \\
    \Omega \dot{\Phi} \label{eq::eomOmegaV} &=& V_N\,.
\end{eqnarray}
We note that the set of equations \eqref{eq::eomInf} just reduce to two simple equations. The first, eq.~\eqref{eq::eomPHIdotV}, has the form of the equation for a single scalar field in a FLRW spacetime. The second, eq.~\eqref{eq::eomOmegaV}, involves the turning rate $\Omega$ and it is not affected by the friction term. Since only the first equation is altered, we can draw the same conclusions about the case $D_t T^a = 0$, as discussed in the previous sub-section. In appendix \ref{appA}, we show that  the friction term can be nevertheless eliminated by an appropriate affine reparametrisation.

\subsection{Multi-field cosmic acceleration}\label{SecMF}

As final step, we consider the coupled system and include the backreaction of the scalar dynamics on the FLRW background. We will explicitly state the conditions required to achieve cosmic acceleration. The action we consider is as before, eq.~\eqref{eq::actionFRLW}. Therefore, the background dynamics of the full system is given by
\begin{eqnarray}
	3 H^2 - \frac{1}{2} \dot{\Phi}^2 - V &=&0\,,\label{eq::FriedmannEQ}\\
	 \ddot{\Phi} + 3H \dot{\Phi}+ V_T &=& 0\,, \label{eq::eomPHIdotV2} \\
    \Omega \dot{\Phi}  &=& V_N\,.
\end{eqnarray}
where the first equation is the Friedmann equation associated to the FLRW metric while the last two equations refer to the dynamics of the scalar fields and are already in projected form, as introduced before.

Cosmic acceleration happens when $\ddot{a}>0$. One can show that this is equivalent to require
\begin{equation}
    \epsilon < 1\,,
\end{equation}
with $\epsilon$ being the {\it acceleration parameter} defined and equal to
\begin{equation}
\epsilon \equiv - \frac{\dot{H}}{H^2} = \frac{\dot{\Phi}^2}{2 H^2} \,.\label{eq::epsilon}
\end{equation}
To require that the acceleration phase lasts for a sufficient number of Hubble times\footnote{This condition is particularly relevant in the case of cosmic inflation to solve the horizon problem. In this case, it is necessary for $\epsilon$ to remain small for a minimum of around 60 e-foldings. In the case of quintessence dark energy, this condition can be relaxed.}, one can require
\begin{equation}
    \eta \equiv \frac{\dot{\epsilon}}{H \epsilon} = 2\epsilon  + 2 \frac{\ddot{\Phi}}{H\dot{\Phi}}<1\,.\label{eq::eta}
\end{equation}
Note that the latter expressions  are exact and do not assume any slow-roll condition. They can be obtained simply by differentiating eq.~\eqref{eq::FriedmannEQ} with respect to the cosmic time $t$ and combining this with eq.~\eqref{eq::eomPHIdotV2}, once we observe that $V_T\ \dot{\Phi} = T^a V_a \dot{\Phi}=\dot{\Phi}^a V_a = \dot{V}$.

Using these definitions, one can rewrite the Friedmann equation \eqref{eq::FriedmannEQ} simply as
\begin{equation}
	H^2=\frac{V}{3-\epsilon}\,.
\end{equation}
Finally, one can derive an expression, which relates the fractional gradient of $V$ and the acceleration parameter $\epsilon$ in a multi-field setup. Let us note that
\begin{equation}
	\frac{|\nabla V|^2}{V^2} = \frac{V_T^2+V_N^2}{V^2}\,,
\end{equation}
where we have used that $V^a=T^a V_T +N^a V_N$. We can obtain an expression for the tangential derivative of $V$ as 
\begin{equation}
	V_T^2=\frac12 \epsilon \left(6-(2\epsilon-\eta)\right)^2H^4\,,
\end{equation}
by combining eq.~\eqref{eq::eomPHIdotV2}, eq.~\eqref{eq::epsilon} and the expression for $\eta$ given in eq.~\eqref{eq::eta}. Similarly, we can obtain an expression for the normal derivative of $V$, namely
\begin{equation}
	V_N^2=2 \Omega^2 H^2 \epsilon\,.
\end{equation}
By combining the last four numbered equations, one finally obtains
\begin{equation}
	\frac{|\nabla V|^2}{V^2}  = 2\epsilon \left( \left( 1+ \frac{\eta}{2(3-\epsilon)} \right)^2 + \frac{\Omega^2}{H^2(3-\epsilon)^2} \right)\,.
\end{equation}
If we demand a phase of cosmic acceleration, namely $\epsilon<1$, then one has
\begin{equation}
	\frac{|\nabla V|^2}{V^2}  \simeq 2\epsilon \left( \left( 1+ \frac{\eta}{6} \right)^2 + \frac{\Omega^2}{9 H^2} \right)\,.\label{eq::gradientVNSR}
\end{equation}
The latter expression shows that one may full-fill the de Sitter conjecture \cite{Obied:2018sgi}, in an accelerating background,  either by having a large turning rate $\Omega$ (namely, a misalignment between the tangent vector and the gradient flow of $V$) or a large $\eta$ parameter (see \cite{Tasinato:2023ukp} for a recent analysis of this regime in the context of single field inflation). If one instead insists on $\eta\ll1$, then one effectively requires a slow roll condition, namely $\ddot{\Phi}\ll H\dot{\Phi}$. In this regime, one obtains
\begin{equation}
	\frac{|\nabla V|^2}{V^2} \simeq  2\epsilon \left( 1+  \frac{\Omega^2}{9H^2} \right)\,,\label{eq::gradientVSR}
\end{equation}
which was already displayed in the introduction section of this work as eq.~\eqref{MFepsilon}.

\section{Asymptotic acceleration and bound on the turning rate}
\label{sec::Omega}

The boundary of moduli space provides an ideal testing ground to examine the predictions of the SDC. It allows for trajectories that extend infinitely, enabling the identification of a tower of states with exponentially decreasing mass along such paths.\footnote{The Emergence String Conjecture \cite{Lee:2019wij} implies that the tower can be represented by either Kaluza-Klein modes or tensionless strings. However, for the purposes of our discussion, the specific nature is not relevant.} Around these asymptotic regions, effective field theories exhibit significant simplifications and possess distinct features.  Recent investigations \cite{Calderon-Infante:2022nxb,Shiu:2023nph,Shiu:2023rxt} have therefore focused on studying cosmic acceleration in these limits.

In this section, we examine the implications of the SDC on a multi-field system that leads to `asymptotic acceleration', referring to cosmic acceleration occurring at the boundary of moduli space. We focus on {\it infinite-distance trajectories}, namely paths in field space that can approach such asymptotic regions. These trajectories can either follow geodesics or deviate from them by a certain angle $\theta$, as already discussed in sec.~\ref{sec:SDCmoduli}. The SDC imposes a strict constraint on this deviation angle, requiring it to approach a constant value in the full infinite-distance limit \cite{Calderon-Infante:2020dhm}. In this section, we specifically consider trajectories with a constant deviation angle from a geodesic throughout the duration of the acceleration.\footnote{Moving away from the boundary allows to have more freedom, such as path-dependent deviations from geodesic trajectories. We will consider this case in the following section.} Moreover, we focus on {\it hyperbolic spaces}, as prototype geometries of infinite distance limits of Calabi–Yau compactifications.

As a key result, we find that the turning rate of such infinite-distance trajectories is negligible, during the acceleration phase. It takes, in fact,  the following general form:
\begin{equation}\label{eq::OmegaBound2}
    \frac{\Omega}{H}=F(\theta, R) \ \sqrt{\epsilon}\,,
\end{equation}
with $F$ being a function of the deviation angle $\theta$ and proportional to the (sectional) curvature of the field manifold.   We also show that this function $F$ is upper bounded by an order one quantity such as
\begin{equation}\label{eq::OmegaBound3}
    F(\theta, R) <F(\theta_0, R)=\mathcal{O}(1)\,,
\end{equation}
where $\theta_0$ is the maximum allowed deviation angle, which is related to the lowest allowed value $\lambda_0$ of the tower mass decay rate (as shown in eq.~\eqref{eq::deviationangle}). The precise form of the function $F$ depends on the specific case and dimensionality of the hyperbolic space and the class of trajectories being followed. It is worth noting that going beyond eq.~\eqref{eq::OmegaBound3} and allowing for larger turning rates would require either a very high curvature of the internal space (see also \cite{Aragam:2021scu}) or considering a product of an unnaturally large number of hyperbolic spaces (see sec.~\ref{sec:NHyp}). However, we argue that this is not a typical situation in generic string effective theories.

 We will proceed as follows. First, in sec.~\ref{sec:oneHyp}, we begin by considering the simplest case of a single hyperbolic plane, which corresponds to a typical axion-saxion system.
 Next, in sec.~\ref{sec:twoHyp}, we move on to a more complex scenario by considering a product of two hyperbolic planes. We will explore the diverse trajectory possibilities that arise in this setup. Finally, in sec.~\ref{sec:NHyp}, we extend our analysis to the case of $N$ hyperbolic planes and generalize our derived formulas.

\subsection{One hyperbolic plane}\label{sec:oneHyp}

Let us begin with a system of two real scalar fields $\Phi^a = (s,\phi)$, namely the saxion $s$ and the axion $\phi$. Their kinetic term is such that it defines an internal field space with hyperbolic geometry. The metric of a single hyperbolic upper half-plane is given by
\begin{equation}
    \rmd\Delta^2 = G_{ab}\ \rmd\Phi^a \rmd\Phi^b= \frac{n^2}{s^2} \left( \rmd s^2 + \rmd\phi^2\right)\,, \label{eq::MetricSinglePlane}
\end{equation}
with $n>0$ controlling the curvature of the field manifold, which reads $R=-2/n^2$. 

Infinite-distance geodesics of this field space are those characterised by constant value of the axion $\phi$ and extend to infinity in the $s$-direction. Along these geodesics, we assume that the mass of an infinite tower of states decreases as $m_s \sim s^{-a}$ for some positive constant $a$.  Other geodesics are semicircles but, in fact, they explore just finite regions of the moduli space and, therefore, will not be considered for our purposes. To test the SDC, we are interested just in the region of large $s$.\footnote{As it was already pointed out in \cite{Scalisi:2018eaz,Calderon-Infante:2020dhm}, the duality of the system under $s\to 1/s$ is just an artifact of this simple model. In realistic string frameworks, this duality is in fact broken once we move away from the boundary and include corrections.}

\begin{figure}[t]
	\begin{center}
		\includegraphics[scale=0.47]{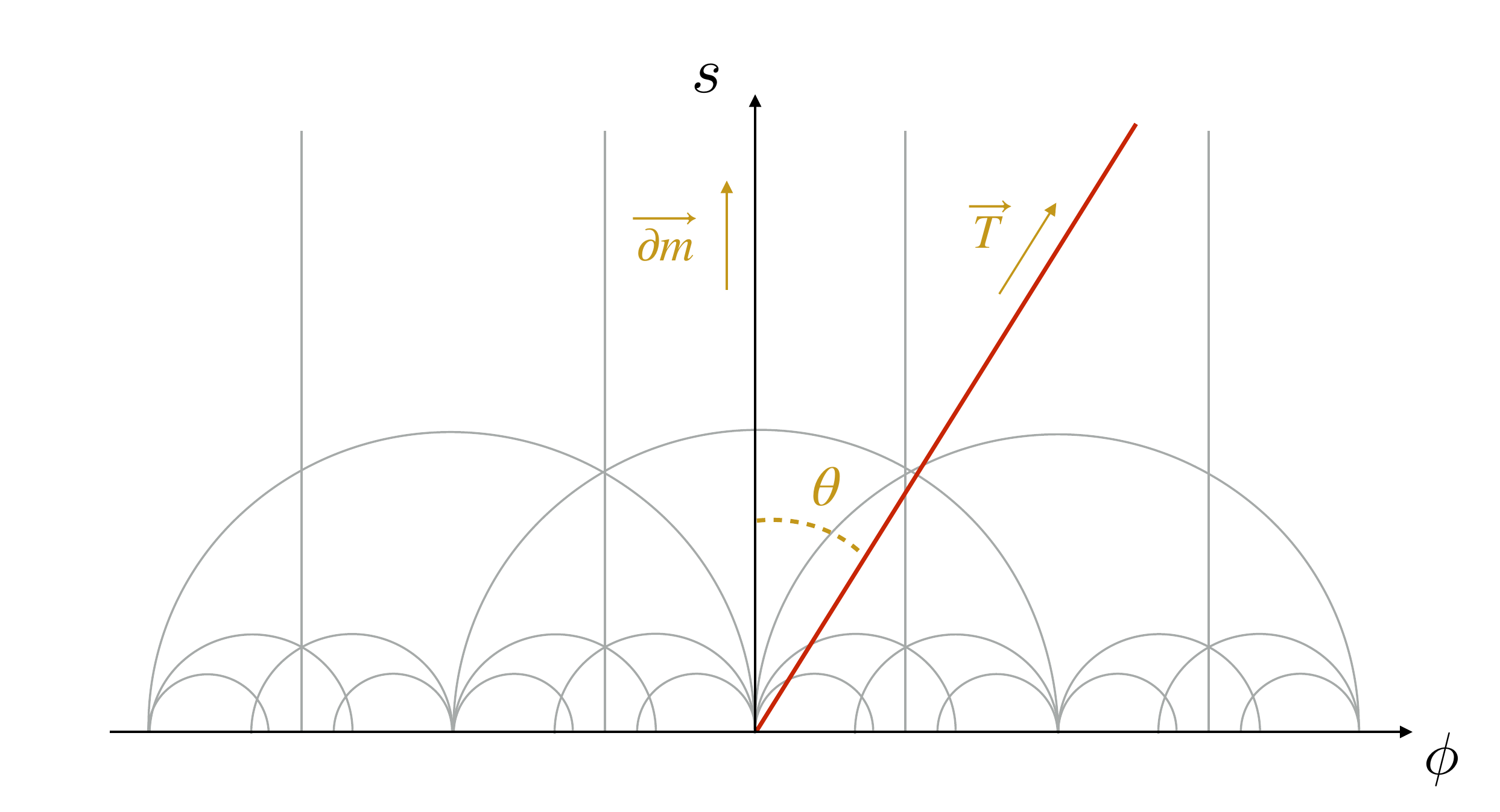}
		\vspace*{-0.2cm}\caption{\it Geodesics of one hyperbolic plane. Just the set of geodesics with constant $\phi$ can approach the infinite distant region. The red line represents a trajectory with a constant deviation angle $\theta$ from the infinite-distance geodesic.}
		\label{Constant}
	\end{center}
\end{figure}

We now consider a family of trajectories deviating by a constant angle $\theta$ from the infinite-distance geodesics (parallel to the $s$-axis) and then apply what we have learned in sec.~\ref{sec:Trajctories} to calculate the turning rate (see fig.~\ref{Constant}). These paths have a non-zero velocity in both axionic and saxionic directions. For our convenience, we define the ratio of these velocities as
\begin{equation}
   \beta\equiv\frac{\dot{\phi}}{\dot{s}} = \frac{\rmd\phi}{\rmd s} \,,\label{eq::critcase1}
\end{equation}
and we will show that this is constant for this class of trajectories. Constant deviations from geodesics are characterized by
\begin{equation}
   G_{ab}\ T_g^a T^b =  \cos\theta = \text{const}\,,
\end{equation}
where $T_g^a$ is the  unit vector tangent to the geodesic trajectory and $T^a$ is the unit vector tangent to the trajectory followed in field space. The components of $T_g^a$ read 
\begin{equation}
    T_g^s = \frac{1}{\dot{\Phi}_g} \dot{s}\,,\qquad
    T_g^\phi = 0\,,\qquad\text{with}\qquad\dot{\Phi}_g = n \frac{\dot{s}}{s}\,,
\end{equation}
while the components of $T^a$ read 
 \begin{equation}
    T^s = \frac{1}{\dot{\Phi}} \dot{s}\,,\qquad
    T^\phi = \frac{1}{\dot{\Phi}} \dot{\phi}= \frac{1}{\dot{\Phi}} \beta\dot{s}\,,
\end{equation}
with the speed along the trajectory given by
\begin{equation}
    \dot{\Phi} = \frac{n}{s} \sqrt{\dot{s}^2 + \dot{\phi}^2} = n \frac{\dot{s}}{s}\sqrt{1+\beta^2} \,.
\end{equation}
Given the above equations, one obtains that 
\begin{equation}
 \cos\theta = \frac{1}{\sqrt{1+\beta^2}}\,,
\end{equation}
which means that trajectories, with a constant deviation angle $\theta$ from a geodesic, have also a constant ratio between the velocities $\dot\phi$ and $\dot{s}$ such that
\begin{equation}
    \beta=\frac{\rmd\phi}{\rmd s} = \tan\theta=\text{const}\,.\label{eq::critcase1} 
\end{equation}

We can now calculate the turning rate in this setup. This is given by
\begin{equation}
    \Omega = \frac{n}{s} \sqrt{ (D_t T^s)^2 + (D_t T^\phi)^2 }\,,\label{eq::Omege1Hyp}
\end{equation}
with
\begin{eqnarray}
    D_t T^s &=& \dot{T}^s + \Gamma_{bc}^s T^b \dot{\Phi}^c = \dot{T}^s - \frac{1}{s} T^s \dot{s} + \frac{1}{s} T^\phi \dot{\phi}\,,\\
    D_t T^\phi &=&  \dot{T}^\phi + \Gamma_{bc}^\phi T^b \dot{\Phi}^c = \dot{T}^\phi - \frac{1}{s} T^s \dot{\phi} - \frac{1}{s} T^\phi \dot{s}\,,
\end{eqnarray}
where we have used the fact that, for the single hyperbolic plane, the only non-vanishing Christoffel symbols are $ \Gamma_{ss}^s =\Gamma_{s\phi}^\phi=-1/s$ and $\Gamma_{\phi\phi}^s
    =1/s$.
For trajectories with a constant deviation angle, as defined by eq.~\eqref{eq::critcase1}, we get
\begin{eqnarray}
    D_t T^s &=&  \frac{1}{\dot{\Phi}} \left(  \ddot{s} - \frac{\dot{s}}{\dot{\Phi}} \ddot{\Phi} -\frac{\dot{s}^2}{s} (1-\beta^2) \right)\,,\label{eq::covder1Hyp1} \\
    D_t T^\phi &=& \frac{1}{\dot{\Phi}} \beta  \left(  \ddot{s} - \frac{\dot{s}}{\dot{\Phi}} \ddot{\Phi} -2 \frac{\dot{s}^2}{s} \right)\,.\label{eq::covder1Hyp2}
\end{eqnarray}
Using the relation $\dot{s}\ddot{\Phi} /\dot{\Phi}=\ddot{s} -\dot{s}^2/s$, the above equations become
\begin{eqnarray}
    D_t T^s &=& \frac{\beta^2}{\dot{\Phi}} \frac{\dot{s}^2}{s} = \frac{\beta^2}{n\sqrt{1+\beta^2}} \dot{s}\,,  \label{DtTs}\\
    D_t T^\phi &=& -\frac{\beta}{\dot{\Phi}} \frac{\dot{s}^2}{s} = -\frac{\beta}{n \sqrt{1+\beta^2}} \dot{s}\,. \label{DtTphi}
\end{eqnarray}

The turning rate then finally reads
\begin{equation}
    \Omega =\beta \frac{\dot{s}}{s} = \frac{\beta}{n\sqrt{1+\beta^2}} \dot{\Phi}\,, \label{eq::OneHypOmegaInBeta}
\end{equation}
namely, the trajectory defined by \eqref{eq::critcase1} is not a geodesic of this field space. As already said, geodesics in the hyperbolic plane are well-known to be just vertical lines and semicircles. More about that can also be found in appendix \ref{appB}. The non-geodesic nature of the trajectory implies that in order to move in a non-vertical straight line on a hyperbolic plane, a normal force $V_N$ needs to be applied. This is indeed an unintuitive property of hyperbolic spaces and one of their unusual characteristics. The presence of the normal force is required to counteract the curvature of the space and allow for straight-line motion.

Finally, we express the speed along the trajectory and the turning rate in terms of the deviation angle as
\begin{eqnarray}
    \dot{\Phi} &=& \frac{n}{\cos\theta} \frac{\dot{s}}{s}\,, \label{eq:speed1Hyptheta} \\
    \Omega &=& \frac{|\sin\theta|}{n} \dot{\Phi}\,. \label{eq::OneHypOmega}
\end{eqnarray}
Assuming cosmic acceleration along this non-geodesic trajectory, we can use eq.~\eqref{eq::epsilon}, which relates the speed along the path to the acceleration parameter, and finally obtain
\begin{equation}\label{eq::Omega1Hyp}
    \frac{\Omega}{H} = \frac{|\sin\theta|}{n} \sqrt{2\epsilon}\,.
\end{equation}
Recalling that the field space curvature is given by $R=-2/n^2$, the above equation can be written in the form as given in the introduction of this section, namely
\begin{equation}
    \frac{\Omega}{H} =F(\theta,R)\sqrt{\epsilon}\,,\qquad\text{with} \qquad F(\theta,R)= |\sin\theta|\sqrt{-R}\,.
\end{equation}
The function $F$ is upper bounded as  $F< |\sin\theta_0| \sqrt{-R}$, with $\theta_0$ being the maximum possible value allowed by the SDC, as described in sec.~\ref{sec:SDCmoduli}. This upper bound is typically an order one quantity. Just unnaturally big curvatures would allow to obtain large turning rates in this framework (a similar conclusion was reached by \cite{Aragam:2021scu}). Using eq.~\eqref{eq::deviationangle}, we can express the bound in terms of the decay rate of the tower mass. We then obtain that the turning rate is directly bounded by 
\begin{equation}
    \frac{\Omega}{H} <\frac{\sqrt{ \lambda_0^2-\lambda_g^2}}{\lambda_g}\sqrt{ R\ \epsilon}\,,\label{eq::boundOmegalambda}
\end{equation}
where we recall that $\lambda_g$ is the decay rate along the geodesic path, while we use $\lambda_0$ to generically mean a lower bound as imposed by string theory (see \cite{Andriot:2020lea,Gendler:2020dfp,Lanza:2021udy,Castellano:2021yye,Etheredge:2022opl} for some works in this direction).

The result derived in equation \eqref{eq::Omega1Hyp} indicates that the turning rate of the trajectory is severely constrained, being proportional to the small parameter $\sqrt{\epsilon}$. This implies that  asymptotic acceleration is predominantly geodesic, with deviations from the geodesic path being highly suppressed. We have obtained this result within the framework of a single hyperbolic plane. In the next section, we will further investigate more complex setups to explore the behavior of trajectories in those cases.

\subsection{Two hyperbolic planes}\label{sec:twoHyp}

As soon as we include a second hyperbolic plane, the situation becomes more intricate. In a product of two hyperbolic planes, we have four real scalar fields $\Phi^a=(s,\phi,u,\psi)$, namely 2 saxions $s,u$ and 2 axions $\phi,\psi$. In this case, the metric takes the following form:
\begin{equation}
    \rmd\Delta^2 = G_{ab}\ \rmd\Phi^a \rmd\Phi^b= \frac{n^2}{s^2} \left( \rmd s^2 + \rmd\phi^2\right) +\frac{m^2}{u^2} \left( \rmd u^2 + \rmd\psi^2\right)\,.
\end{equation}
 We associate a tower of states with each saxionic direction such that the respective masses decrease as  $m_s \sim s^{-a}$ and $m_u \sim u^{-b}$, with $a$ and $b$ some positive constants. Beside the purely saxionic directions, we now have a wider range of possibilities for infinite-distance trajectories. A trivial choice is to keep  $u$ and $\psi$ fixed (or $s$ and $\phi$ fixed). In this case, we immediately recover results analogous to the single hyperbolic plane. In the following, instead, we consider other possible non-trivial cases.

 \subsubsection{Saxion-axion trajectories}\label{sec:axionsaxion2Hyp}
As a first non-trivial case, we consider trajectories with $u$ and $\phi$ fixed ($u=u_0$ and $\phi=\phi_0$). Namely, we consider the saxion $s$, from the first hyperbolic plane, evolving together with the axion $\psi$, from the second hyperbolic plane. In particular, analogously to sec.~\ref{sec:oneHyp}, we consider paths deviating by a constant angle from the infinite-distance geodesics (namely trajectories for which just the saxion $s$ evolves). This, again, corresponds to the case of constant ratio of velocities along the two directions, namely
\begin{equation}
   \gamma\equiv s \frac{\dot{\psi}}{\dot{s}}= \text{const}\,. \label{eq::SaxAxRel2}
\end{equation}
Notice that, unlike in the previous section, the formula above contains a factor of $s$. This arises because the field $u$, which multiplies the kinetic term of $\psi$, has been set to a constant value, while the saxion $s$ retains its non-canonical kinetic term. In section \ref{sec:oneHyp}, both the fields $s$ and $\phi$ had kinetic terms with a multiplying factor of $1/s$, which canceled out in the ratio given by eq.~\eqref{eq::critcase1}. One can now calculate the deviation angle from the geodesic trajectory and obtain
\begin{equation}
    \cos\theta = G_{ab}\ T_g^a T^b = \left(1 + \frac{m^2\gamma^2}{u_0^2 n^2}\right)^{-\frac12}\,,\label{eq::RiemAngleDef}
\end{equation}
which gives the identification
\begin{equation}
\tan\theta = \frac{m\gamma}{u_0 n}\,.
\end{equation}
From the equations above, we can deduce that the constant ratio of velocities $\gamma$ corresponds in fact to the constant deviation angle $\theta$ from a geodesic trajectory.

We can now calculate the turning rate in this setup. This is given by
\begin{equation}
    \Omega^2 = \frac{n^2}{s^2} \left( (D_t T^s)^2 + (D_t T^\phi)^2 \right) + \frac{m^2}{u^2} \left( (D_t T^u)^2 + (D_t T^\psi)^2 \right)\,.
\end{equation}
After some algebra, the covariant derivatives turn out to be 
\begin{equation}
    D_t T^u = \frac{1}{u_0} \frac{1}{\dot{\Phi}} \dot{\psi}^2 = \frac{1}{u_0} \frac{1}{\dot{\Phi}} \gamma^2 \frac{\dot{s}^2}{s^2}\,, \qquad D_t T^s=D_t T^\phi=D_t T^\psi=0\,, \label{eq::DtTuSaxionAxion}
     \end{equation}
with the speed along the trajectory given by
\begin{equation}
    \dot{\Phi}^2 = \frac{n^2}{s^2} \dot{s}^2 + \frac{m^2}{u_0^2} \dot{\psi}^2 = n^2 \frac{\dot{s}^2}{s^2} \left(1+\frac{m^2\gamma^2}{u_0^2n^2}\right)\,.
\end{equation}
Eq.~\eqref{eq::DtTuSaxionAxion} shows that, despite the fact that the saxion $u$ is taken to be constant, the only non-zero contribution to the turning rate is given by the projection of the covariant derivative in the $u$-direction. Namely, for this class of trajectories, the tangent vector fails to be parallel transported along the $u$-direction. We can then express the turning rate as
\begin{equation}
    \Omega^2 = \frac{m^2 \gamma^4}{u_0^4} \frac{1}{\dot{\Phi}^2} \left( \frac{\dot{s}^2}{s^2}\right)^2 = \frac{\frac{m^2\gamma^4}{u_0^4n^4}}{\left( 1 + \frac{m^2\gamma^2}{u_0^2n^2} \right)^2} \dot{\Phi}^2\,,
\end{equation}
where we immediately recognise the same quadratic dependence of the speed $\dot{\Phi}$ as in eq.~\eqref{eq::OneHypOmega}. The speed and the turning rate can be now given in terms of the deviation angle as
\begin{eqnarray}
    \dot{\Phi} &=& \frac{n}{\cos\theta} \frac{\dot{s}}{s}\,, \\
    \Omega &=& \frac{\sin^2\theta}{m} \dot{\Phi}\,.
\end{eqnarray}
The speed $\dot{\Phi}$ has formally the same expression as for the single hyperbolic plane. However, the turning rate $\Omega$ has instead a different power of the sine, when compared to eq.~\eqref{eq::OneHypOmega}. Moreover, the curvature parameter appearing in $\Omega$ is the one of the second hyperbolic plane, reflecting the fact that the tangent vector turns with respect to the saxion $u$. 

Finally, assuming cosmic acceleration along this non-geodesic trajectory, and using eq.~\eqref{eq::epsilon} to relate the speed to the acceleration parameter, we obtain
\begin{equation}
    \frac{\Omega}{H} = \frac{\sin^2\theta}{m} \sqrt{2\epsilon}.
\end{equation}
One can draw conclusions analogous to those presented for the case of a single hyperbolic plane. At the boundary of this moduli space, asymptotic acceleration is mainly geodesic. The presence of a non-geodesic trajectory introduces deviations from geodesic behavior, but the turning rate remains small compared to the Hubble parameter, indicating that geodesic motion dominates the dynamics.

 \subsubsection{Saxion-saxion trajectories}\label{sec:saxionsaxion2Hyp}

Here we consider another two-field infinite-distance trajectory where we keep the two axions $\phi$ and $\psi$ fixed and allow the two saxions $s$ and $u$ to evolve together. Again, this class of trajectories is characterized by a constant ratio of the velocities along the two directions, namely
\begin{equation}
    \delta\equiv\frac{\dot{u}}{\dot{s}}\frac{s}{u}= \text{const}\,, \label{eq::SaxSaxRel}
\end{equation}
where we have considered that both saxions have non-canonical kinetic terms. With this definition, the speed along the trajectory becomes
\begin{equation}
    \dot{\Phi}^2 = \frac{n^2}{s^2} \dot{s}^2 + \frac{m^2}{u^2} \dot{u}^2 = n^2 \frac{\dot{s}^2}{s^2} \left(1+\frac{m^2}{n^2}\delta^2 \right)\,,
\end{equation}
while the tangent vector components read
\begin{equation}
    T^s = \frac{1}{\dot{\Phi}} \dot{s}\,,\qquad T^u = \frac{1}{\dot{\Phi}} \dot{u} = \frac{u}{\dot{\Phi}} \delta \frac{\dot{s}}{s}\,,\qquad
    T^\phi =T^\psi = 0\,.
\end{equation}
After some algebra, one can prove that 
\begin{equation}
D_t T^s = D_t T^\phi =D_t T^u = D_t T^\psi =  0\,,
\end{equation}
which directly implies
\begin{equation}
    \Omega = 0\,.
\end{equation}
Hence, a trajectory involving only two saxions is always geodesic for any $\delta$. More about this can be found in appendix \ref{appB}.

\subsubsection{Saxion-axion-axion trajectories}

Here we consider one final combination where only the second saxion $u$ is constant, i.e. $u=u_0$. This is the first trajectory involving the evolution of three fields, which brings another new feature with it. We consider trajectories where the velocities along the three directions satisfy
\begin{equation}
    \left( \frac{\dot{\phi}}{\dot{s}} \right)^2 + \frac{m^2}{n^2 u_0^2} \left(  \frac{\dot{\psi}}{\frac{\dot{s}}{s}} \right)^2 = \text{const}\,.
\end{equation}
However, we do not study this case in full generality, rather we assume that both terms are separately constant, namely
\begin{equation}
     \beta\equiv\frac{\dot{\phi}}{\dot{s}} =\text{const}\,,\qquad
    \gamma\equiv s\frac{\dot{\psi}}{\dot{s}} = \text{const}\,.
\end{equation}
With these definitions, the speed along the trajectory is then given by
\begin{equation}
    \dot{\Phi}^2 = \frac{n^2}{s^2} (\dot{s}^2 +\dot{\phi}^2) + \frac{m^2}{u_0^2} \dot{\psi}^2 = n^2 \frac{\dot{s}^2}{s^2} \left(1+ \beta^2 +\frac{m^2\gamma^2}{u_0^2n^2}\right)
\end{equation}
while the tangent vector components are
\begin{equation}
    T^s = \frac{1}{\dot{\Phi}} \dot{s}\,,\qquad T^\phi = \frac{1}{\dot{\Phi}} \dot{\phi}=\frac{1}{\dot{\Phi}}\beta \dot{s}\,,\qquad
    T^u = 0\,,\qquad
    T^\psi = \frac{1}{\dot{\Phi}} \dot{\psi} = \frac{1}{\dot{\Phi}} \gamma \frac{\dot{s}}{s}\,.
\end{equation}
After some algebra, one obtains that the covariant derivative components are given by
\begin{equation}
    D_t T^s = \frac{\beta^2}{\dot{\Phi}} \frac{\dot{s}^2}{s}\,,\qquad
    D_t T^\phi =  -\frac{\beta}{\dot{\Phi}} \frac{\dot{s}^2}{s}\,,\qquad
    D_t T^u = \frac{1}{u_0} \frac{1}{\dot{\Phi}} \gamma^2 \frac{\dot{s}^2}{s^2}\,,\qquad
    D_t T^\psi = 0\,,
\end{equation}
which implies that the turning rate is
\begin{equation}
    \Omega^2 = \frac{\dot{\Phi}^2}{\left(1+ \beta^2 +\frac{m^2\gamma^2}{u_0^2n^2}\right)^2} \left( \frac{\beta^2}{n^2} (1+\beta^2) + \frac{m^2\gamma^4}{u_0^4n^4} \right).
\end{equation}
Setting either $\beta =0$ or $\gamma =0$ yields the previous cases, which serves as a nice consistency check. Importantly, we observe that the turning rate $\Omega$ still scales with $\dot{\Phi}$, despite the different expressions for the speed $\dot{\Phi}$ in each case, considered so far. This emerges as a universal feature and it proves to be a crucial property when relating the results to cosmic acceleration.

We can again calculate the angle between the tangent vector to the geodesic (any line parallel to the $s$-axis) and the tangent vector to the trajectory. Thus, we have 
\begin{equation}
    \cos\theta = G_{ab}\ T_g^a T^b = \left(1+ \beta^2 +\frac{m^2\gamma^2}{u_0^2n^2}\right)^{-\frac12}\,,
\end{equation}
such that
\begin{equation}
    \tan\theta = \sqrt{\beta^2 +\frac{m^2\gamma^2}{u_0^2n^2}}\,.
\end{equation}
Furthermore, we can define the angle $\theta_\phi$ in the $s$-$\phi$-plane by setting $\gamma =0$ in $T^a$ and analogously the angle $\theta_\psi$ in the $s$-$\psi$-plane by setting $\beta =0$. We then have
\begin{equation}
     \tan\theta_\phi = \beta\,,\qquad
    \tan\theta_\psi = \frac{m\gamma}{n u_0}\,.
\end{equation}
This enables us to express the speed $\dot\Phi$ and the turning rate $\Omega$ in terms of these angles as
\begin{eqnarray}
    \dot{\Phi}^2 &=& \frac{n^2}{\cos^2\theta} \frac{\dot{s}^2}{s^2}\,, \\
    \Omega^2 &=& \cos^4\theta \left( \frac{1}{n^2} \frac{\tan^2\theta_\phi}{\cos^2\theta_\phi} + \frac{1}{m^2} \tan^4\theta_\psi \right) \dot{\Phi}^2\,.
\end{eqnarray}
In this formulation, the connection to the previous results becomes even more apparent. If we set $\gamma =0$, we get $\theta= \theta_\phi$ and $\theta_\psi=0$ thereby recovering the result of the single hyperbolic plane. Of course, the same logic works for setting $\beta =0$.
Finally, if we assume that cosmic acceleration occurs along this trajectory, then we obtain
\begin{equation}
    \frac{\Omega^2}{H^2} = 2\epsilon \cos^4\theta \left( \frac{1}{n^2} \frac{\tan^2\theta_\phi}{\cos^2\theta_\phi} + \frac{1}{m^2} \tan^4\theta_\psi \right)\,,
\end{equation}
where the trigonometric function is bounded to be $\text{max}\{m^{-2}, n^{-2}\}$, namely, an order one factor. This result indicates once more that trajectories leading to asymptotic acceleration must have a negligible turning rate.

\subsection{$N$ hyperbolic planes} \label{sec:NHyp}

In this section, we extend our computations to an arbitrary number $N$ of hyperbolic planes. The generalization to $N$ hyperbolic planes follows the same principles discussed for the cases of one and two hyperbolic planes. Each additional hyperbolic plane introduces more components and equations, but they can be categorized into the two base cases: saxion with an axion from the same hyperbolic plane, and saxion with an axion from another hyperbolic plane. We do not discuss the case of several saxions since we have already seen that it leads to zero contribution to the turning rate $\Omega$.

We consider the product of $N$ hyperbolic planes with $N$ saxions $s_i$ and $N$ axions $\phi_i$, making  a total of $2N$ real scalar fields $\Phi^a = (s_1,\phi_1,...,s_N,\phi_N)$. The metric of this field space is
\begin{equation}
    \rmd\Delta^2 = G_{ab}\ \rmd\Phi^a \rmd\Phi^b= \sum_{i=1}^N \frac{n^2_i}{s^2_i} \left(\rmd s_i^2 + \rmd\phi_i^2\right)\,.
\end{equation}
with $n_i$ being the curvature parameter of each $i$-th hyperbolic plane. Also in this case, we assume that along each saxionic direction a tower of states will have decreasing mass as $m_{s_i} \sim s^{-a_i}$, for some constants $a_i>0$.

From now on, we fix a saxionic direction, without loss of generality. We choose $s_1$ and drop the index 1 from all quantities of the first hyperbolic plane, i.e. $s_1\equiv s$, $\phi_1\equiv \phi$ and $n_1\equiv n$. Next, we fix the trajectory as the one which involves displacement of the saxion $s$ and of all axions $\phi$ and $\phi_i$.  The other saxions are taken to be constant, that is $s_i = \text{const}$ for $i\neq 1$. To simplify the notation, we  drop the label 0 here, so we write $s_i$ instead of $s_{0i}$.
By analogy with the previous subsection, these trajectories satisfy
\begin{equation}
    \left( \frac{\dot{\phi}}{\dot{s}} \right)^2 + \sum_{i=2}^N \frac{n_i^2}{n^2 s_i^2} \left(  \frac{\dot{\phi_i}}{\frac{\dot{s}}{s}} \right)^2 = \text{const}\,.
\end{equation}
Again, we simplify the situation by assuming that all terms are individually constant, such as
\begin{equation}
   \beta\equiv \frac{\dot{\phi}}{\dot{s}}  = \text{const}\,,\qquad
     \beta_i\equiv s\frac{\dot{\phi_i}}{\dot{s}} = \text{const}\,,\qquad \text{for }i\geq 2\,.
\end{equation}
With these definitions, we can write the expression of the speed along the trajectory as
\begin{equation}
    \dot{\Phi}^2 = \frac{n^2}{s^2} (\dot{s}^2 +\dot{\phi}^2) + \sum_{i=2}^{N} \frac{n_i^2}{s_i^2} \dot{\phi_i}^2 = n^2 \frac{\dot{s}^2}{s^2} \left(1+ \beta^2 + \sum_{i=2}^N \frac{n_i^2\beta_i^2}{s_i^2n^2}\right)\,.
\end{equation}
The turning rate thus becomes
\begin{equation}
    \Omega^2 = \frac{\dot{\Phi}^2}{\left(1+ \beta^2 + \sum_{i=2}^N \frac{n_i^2\beta_i^2}{s_i^2n^2} \right)^2} \left( \frac{\beta^2}{n^2} (1+\beta^2) + \sum_{i=2}^N \frac{n_i^2\beta_i^4}{s_i^4n^4} \right)\,,
\end{equation}
where, once more, we confirm the relation such as $\Omega \sim \dot{\Phi}$.

As in the previous cases, it is again possible to define the angle between the geodesic tangent vector $T_g^a$, which corresponds to setting $\beta=\beta_i=0$, and the tangent vector to the trajectory $T^a$. This deviation angle is given by
\begin{equation}
    \cos\theta = G_{ab}\ T_g^a T^b = \left(1+ \beta^2 + \sum_{i=2}^N \frac{n_i^2\beta_i^2}{s_i^2n^2}\right)^{-\frac12}\,. 
\end{equation}
Furthermore, we can define the angle $\theta_\phi$ in the $s$-$\phi$-plane by setting $\beta_i =0$ for $i\geq2$  and  the angle $\theta_{\phi_i}$ in the $s$-$\phi_i$-plane by setting $\beta =\beta_j=0$ for $j\neq i$. We then have
\begin{equation}
    \tan\theta_\phi = \beta\,,\qquad
    \tan\theta_{\phi_i} =\frac{n_i \beta_i}{n s_i}\,,
\end{equation}
This allows us to express the speed $\dot\Phi$ and the turning rate $\Omega$ in terms of these angles, namely
\begin{eqnarray}
    \dot{\Phi}^2 &=& \frac{n^2}{\cos^2\theta} \frac{\dot{s}^2}{s^2} \\
    \Omega^2 &=& \cos^4\theta \left( \frac{1}{n^2} \frac{\tan^2\theta_\phi}{\cos^2\theta_\phi} + \sum_{i=2}^N \frac{1}{n_i^2} \tan^4\theta_{\phi_i} \right) \dot{\Phi}^2. \label{Ome2}
\end{eqnarray}
This is perfectly consistent with the results of the previous subsections. Finally, if we assume that cosmic acceleration occurs along this path, then one has
\begin{equation}
    \frac{\Omega^2}{H^2} = 2\epsilon \cos^4\theta \left( \frac{1}{n^2} \frac{\tan^2\theta_\phi}{\cos^2\theta_\phi} + \sum_{i=2}^N \frac{1}{n_i^2} \tan^4\theta_{\phi_i} \right)\,,
\end{equation}
where the trigonometric function is upper-bounded by $\text{max}\{n^{-2}, n_1^{-2},\ldots,n_{N-1}^{-2}\}$. This nicely generalizes all previous results.

\section{Moving away from the boundary  of moduli space}
\label{sec::nonconstOmega}

In the previous section, we have demonstrated that the boundary of moduli space highly restricts the possibility of realizing large turning rates in a multi-field setup that leads to cosmic acceleration. We have found this result by focusing on hyperbolic spaces and on trajectories, which have a constant deviation angle from geodesics, namely $\theta=\text{const}$. The SDC, in fact, does not allow for any other non-geodesic behaviour in the full infinite distance limit, as already pointed out in \cite{Calderon-Infante:2020dhm}.

However, the constraints relax when moving away from the boundary, thus allowing trajectories with time-dependent deviations from a geodesic. A time-dependent deviation angle $\theta=\theta(t)$ corresponds to a path-dependent decay rate $\lambda=\lambda(\Delta)$ of the tower, following eq.~\eqref{exprate}. Specifically, one expects a structure like $\lambda=\lambda_\infty + \delta\lambda(\Delta)$, with a leading constant term $\lambda_\infty$ and some corrections $\delta\lambda$ that vanish at the boundary. This leads to corrections to the SDC exponential formula of the tower mass, such as
\begin{equation}
   m =m_0 \exp(-\lambda \Delta) + \delta m (\Delta)
\end{equation}
with $\delta m \rightarrow 0$ in the limit $\Delta\rightarrow \infty$, namely at the boundary.

In this section, we examine the case of a time-dependent deviation angle and explore its implications for the turning rate. We find that achieving a large turning rate $\Omega$ requires  non-generic conditions for the trajectory. After providing the general formulas in sec.~\ref{sec:nonconstanttheta}, we present a specific example of $\theta$ as a Taylor expansion in negative powers of the saxion $s$ in sec.~\ref{sec:asymptheta}. For the sake of simplicity, we focus on the framework of a single hyperbolic plane.

\subsection{Non-constant deviation angle}\label{sec:nonconstanttheta}

\begin{figure}[t]
	\begin{center}
		\includegraphics[scale=0.48]{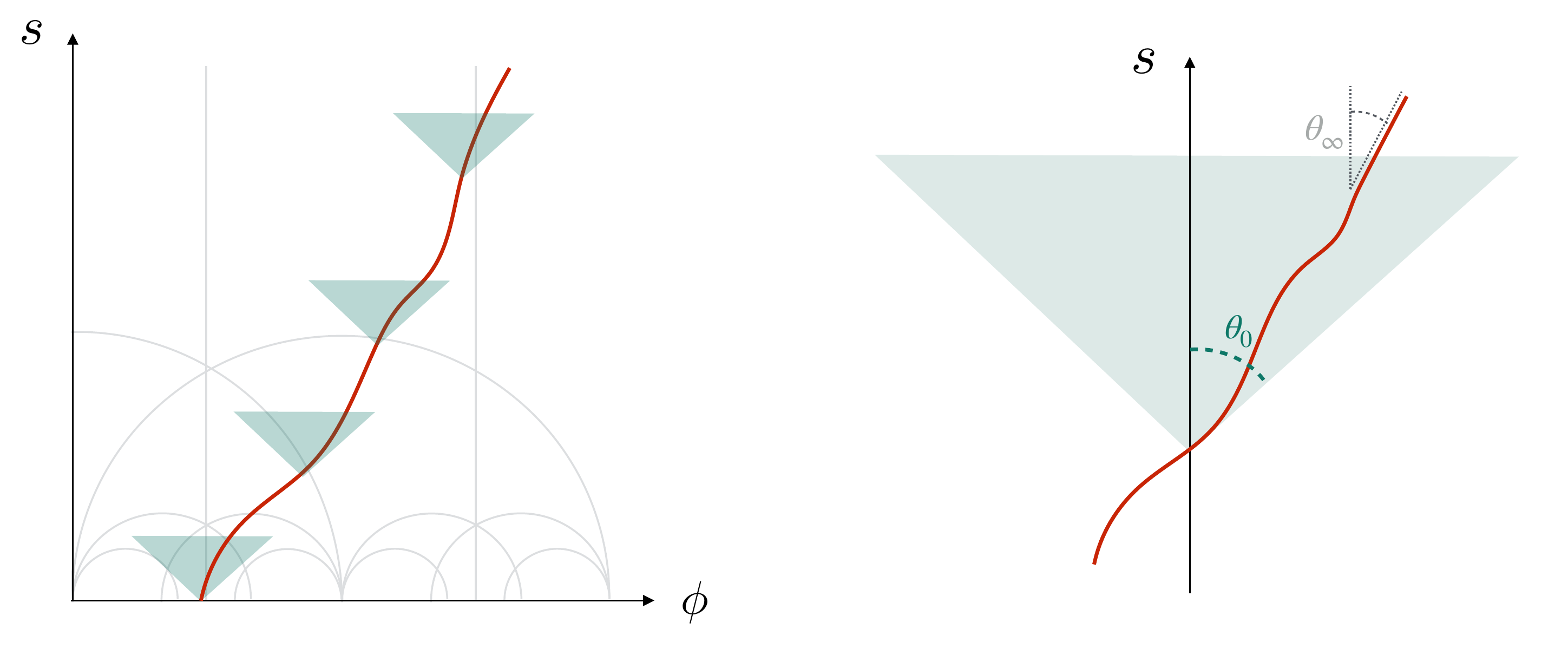}
		\vspace*{-0.2cm}\caption{\it Trajectory with a non-constant deviation angle in a hyperbolic plane. The trajectory, at any point, remains inside the cone $|\theta(s)|\leq \theta_0$, related to the lower bound on the mass decay rate of the tower of states. $\theta_{\infty}$ represents the deviation angle at the boundary ($s\rightarrow\infty$).}
		\label{Non-constant}
	\end{center}
\end{figure}

In the framework of a single hyperbolic plane, considering a non-constant deviation angle corresponds to a time-dependent ratio of the velocities along the saxionic and axionic directions:
\begin{equation}
   \frac{\rmd\phi}{\rmd s}=\beta(t)=\tan\theta(t)\,.
\end{equation}
The trajectory (see fig.~\ref{Non-constant}) is now defined by following tangent vector and speed
\begin{equation}
    T^s = \frac{1}{\dot{\Phi}} \dot{s}\,,\qquad
    T^\phi = \frac{1}{\dot{\Phi}} \beta(t) \dot{s}\,,\qquad
    \dot{\Phi}^2 = n^2 \frac{\dot{s}^2}{s^2} \left( 1+ \beta(t)^2 \right)\,. \label{eq::NonConstTrajTangent}
\end{equation}
Then, we observe all the second derivative terms, which implicitly involve $\beta(t)$, get extra contributions, namely
\begin{eqnarray}
    \frac{\dot{s}}{\dot{\Phi}} \ddot{\Phi} &=& \ddot{s} -\frac{\dot{s}^2}{s} + \frac{\dot{s}}{1+\beta^2} \beta\dot{\beta}\,, \\
    \ddot{\phi} &=& \ddot{s}\beta + \dot{s}\dot{\beta}\,, \\
    \frac{\dot{\phi}}{\dot{\Phi}} \ddot{\Phi} &=& \beta \left( \ddot{s} -\frac{\dot{s}^2}{s} \right) + \frac{\beta^2}{1+\beta^2}\dot{s}\dot{\beta}\,.
\end{eqnarray}
We can plug the above expressions into  eq.~\eqref{eq::covder1Hyp1} and eq.~\eqref{eq::covder1Hyp2} and obtain
\begin{eqnarray}
    D_t T^s &=& \frac{\beta^2}{n\sqrt{1+\beta^2}} \dot{s} - \frac{s \beta}{n\left(1+\beta^2\right)^{\frac{3}{2}}}  \dot{\beta}\,, \\
    D_t T^\phi &=&  -\frac{\beta}{n \sqrt{1+\beta^2}} \dot{s} + \frac{s }{n\left(1+\beta^2\right)^{\frac{3}{2}}}  \dot{\beta}\,,
\end{eqnarray}
which is of course consistent with previous findings upon setting $\dot{\beta}=0$. Switching to the formulation in terms of the deviation angle $\theta$, we obtain
\begin{eqnarray}
    D_t T^s &=& \frac{1}{n} \tan^2\theta \cos\theta ~ \dot{s} - \frac{s}{n} \sin\theta ~ \dot{\theta}\, , \\
    D_t T^\phi &=&  -\frac{1}{n} \sin\theta ~\dot{s} + \frac{s }{n} \cos\theta ~ \dot{\theta} \, ,
\end{eqnarray}
where we used $\dot{\beta}=\dot{\theta}/\cos^2\theta$. Plugging this into the expression of the turning rate eq.~\eqref{eq::Omege1Hyp}, we get
\begin{equation}
    \Omega^2 = \left( \tan\theta\  \frac{\dot{s}}{s} - \dot{\theta} \right)^2 \,.
\end{equation}
Furthermore, using eq.~\eqref{eq:speed1Hyptheta} for the speed $\dot\Phi$ in terms of the angle $\theta$, we arrive at the final result
\begin{equation}
    \Omega = \left\vert  \frac{\sin\theta}{n} \dot{\Phi} - \dot{\theta}  \right\vert\,. \label{eq::nonconstThetaOmega}
\end{equation}
This is still fully consistent with eq.~\eqref{eq::OneHypOmega}, which was obtained in the case of $\dot\theta=0$. In an accelerating background, the first term of the above equation is small, being proportional to $\sqrt{\epsilon}$. This implies that a large turning rate $\Omega$ can be achieved just in the case of large  $\dot{\theta}$. We argue that this is not a generic situation for trajectories, which eventually approach the boundary of moduli space. We give an example in the next subsection.

\subsection{Asymptotic expansion of $\theta$}\label{sec:asymptheta}

Approaching the boundary of moduli space, in the framework of a single hyperbolic space, translates into moving towards large values of $s$. A natural choice of non-constant deviation angle, to parameterize the departure from the boundary, is a Taylor expansion in negative powers of the saxion $s$. Therefore, we consider\footnote{Another option would be to consider an expansion of $\beta(s)=\tan\theta(s)$ in negative powers of the saxion $s$ (this possibility was already suggested in \cite{Scalisi:2018eaz}).}
\begin{equation}
    \theta(s) = \theta_\infty + \sum_{n>0} \frac{c_n}{s^n}  \,,\label{eq::nonconstThetaExpansion}
\end{equation}
where  $\theta_\infty$ is the value of theta at the boundary ($s=\infty$). Note that we still require that
\begin{equation}
    \theta(s)\leq \theta_0\,,\label{eq::thetabound}
\end{equation}
at any point in field space, with $\theta_0$ being the maximum possible value of the angle (see fig.~\ref{Non-constant}). This is to be consistent with the existence of a universal lower bound on the decay rate of the SDC tower mass as, for example, claimed in \cite{Andriot:2020lea,Gendler:2020dfp,Lanza:2021udy,Castellano:2021yye,Etheredge:2022opl}. Let us consider just the constant and the first leading of the above expansion
\begin{equation}
    \theta(s) \simeq \theta_\infty + \frac{c_k}{s^k}\,, \label{eq::approxExpTheta}
\end{equation}
where $k$ must not be necessarily equal to 1. Using eq.~\eqref{eq::thetabound}, we then find 
\begin{equation}
    \frac{c_k}{s^k} \leq \theta_0 - \theta_\infty \leq 2 \theta_0\,,\label{eq::nonconstThetaBound}
\end{equation}
because $\theta_\infty$ could at most be equal to $-\theta_0$. We can now calculate the angular velocity, that is
\begin{equation}
    \dot{\theta}(s) \approx -k  \frac{c_k}{s^k} \frac{\dot{s}}{s}\,.\label{eq::thetabound2}
\end{equation}
We note that, for this type of trajectories, also $\dot\theta$ vanishes in the limit $s\rightarrow \infty$. More importantly, using eq.~\eqref{eq:speed1Hyptheta}  and \eqref{eq::thetabound2}, we can get an upper bound for the absolute value of the angular velocity
\begin{equation}
    \left\vert \dot{\theta}(s) \right\vert \leq \frac{2k}{n} \theta_0 \cos\theta(s) ~ \dot{\Phi}\,.
\end{equation}
If cosmic acceleration occurs along this non-geodesic trajectory, then one can use eq.~\eqref{eq::epsilon} and express the speed $\dot\Phi$ in terms of the Hubble parameter $H$ and the acceleration parameter $\epsilon$. The bound on the angular velocity thus becomes
\begin{equation}
    \left\vert \frac{\dot{\theta}(s)}{H} \right\vert \leq 2\sqrt{2}\ \frac{k}{n}\ \theta_0\ \sqrt{\epsilon}\,,
\end{equation}
namely, also the angular velocity (in Hubble units) is constrained by the parameter $\epsilon$, which is less than unity during acceleration. This result, combined with the more general formula \eqref{eq::nonconstThetaOmega} on the turning rate, again implies
\begin{equation}
    \frac{\Omega}{H} \simeq \sqrt{\epsilon}.
\end{equation}
The significance of this last equation is that, even when moving away from the boundary, the SDC imposes strict constraints on the turning rate of non-geodesic trajectories in an accelerating background.

\section{Conclusions}
\label{sec:concl}

In this work, we have studied the constraints imposed by SDC on multi-field acceleration scenarios at the boundary of moduli space. This is a natural framework for string effective models, which typically involve a rich spectrum of massless and/or light scalar fields. Furthermore, the EFTs  offer simplified perturbative descriptions in these asymptotic regions of the moduli space, making the boundary an ideal setting to extract robust predictions.

As a key result of this investigation, we have found that, in accelerating backgrounds, field trajectories that extend infinitely and satisfy the SDC must exhibit a negligible turning rate. Specifically, we have established that the turning rate $\Omega$ (measured in Hubble units) must be proportional to $\sqrt{\epsilon}$, where $\epsilon$ denotes the acceleration parameter. Since $\epsilon<1$ in a quasi-de Sitter space, this proportionality implies that $\Omega$ is indeed small.
Furthermore, we have shown that the turning rate $\Omega$ is bounded above by a function of the minimum value allowed for the mass decay rate of the SDC tower within string theory \cite{Andriot:2020lea,Gendler:2020dfp,Lanza:2021udy,Castellano:2021yye,Etheredge:2022opl}. A specific expression for this bound can be found in eq.~\eqref{eq::boundOmegalambda}.
We have obtained this result in the context of hyperbolic spaces of different dimensionality (namely, with an arbitrary number of fields), as systematically described in sec.~\ref{sec::Omega}. Furthermore, we have tested this result both for trajectories with a constant deviation from a geodesic (sec.~\ref{sec::Omega}) and with a time-dependent deviation from the geodesics (sec.~\ref{sec::nonconstOmega}). The latter case becomes a convenient way to parameterize departures from the boundary.

This finding aligns perfectly with the well-established understanding that the physics at the boundary of the moduli space is subject to stringent constraints. Several properties characterize this limit. For instance, the species scale $\Lambda_s$ tends to zero (see e.g. \cite{vandeHeisteeg:2023ubh,Cribiori:2023ffn}), while corrections to the K\"ahler- and super-potential of the EFT vanish. The scalar potential  and the gravitino mass approach zero value \cite{Ooguri:2018wrx,Hebecker:2018vxz,Cribiori:2021gbf,Castellano:2021yye}, while  the entropy instead increases (see e.g. \cite{Cribiori:2023ffn}). Our work adds to this list of properties by showing that asymptotic acceleration is primarily geodesic. 

The implications of this result are manifold:  \begin{itemize}

    \item Fulfilling the de Sitter conjecture \cite{Obied:2018sgi,Ooguri:2018wrx}, in an accelerating  background and at the boundary of moduli space, becomes challenging, given our result. Despite the fact that we are in a multi-field setup, the gradient of the scalar potential is predominantly  determined by the acceleration parameter $\epsilon$ (see eq.~\eqref{eq::gradientVSR}), giving results analogous to the single-field case . As pointed out in sec.~\ref{sec:Trajctories}, an alternative interesting way to obtain large values of $|\nabla V|/V$, and satisfy the de Sitter conjecture, is to assume cosmological phases with large parameter $\eta$ (see \cite{Tasinato:2023ukp} for a recent analysis of this situation in single-field inflation). In this case, a slow-roll condition is violated, namely, the second derivatives of the fields are not smaller than the friction Hubble term in the equations of motion.
    
     \item Models engineering cosmic acceleration with rapid turns face significant constraints at the boundary of the moduli space, making it more favorable to explore such scenarios in the \textit{bulk} of the moduli space. In the bulk, both perturbative and non-perturbative corrections are expected to play a fundamental role in shaping the dynamics of the system. These corrections can potentially provide additional degrees of freedom and interactions that allow for more flexible and diverse trajectories, facilitating the realization of desired acceleration patterns.

    \item  The total field displacement $\Delta$ is subject to a universal upper bound $\Delta\lesssim-\log H$, as established in \cite{Scalisi:2018eaz}. In the context of multi-field inflation, our findings suggest that the total field displacement remains bounded by the tensor-to-scalar ratio measured at CMB scales, denoted as $\Delta\lesssim-\log r$. Although the original derivation of this bound in \cite{Scalisi:2018eaz} was focused on single-field scenarios, it is applicable to the case of multi-field inflation due to the fact that trajectories must be (quasi-)geodesic.

    \item Our result on the turning rate, in the context of inflation, implies also that the speed of sound $c_s$ of primordial perturbations must be close to unity, as in the single-field case. The expression of the speed of sound in multi-field models is indeed $c_s=\left(1+4\Omega^2/{M^2}\right)^{-1/2}$, with $M$ being the mass of the fluctuations orthogonal to the trajectory and typically assumed larger than the Hubble scale $H$. This seems to be consistent with the results of \cite{Hetz:2016ics}, which found a stringent lower bound on $c_s$ in the context of $\mathcal{N}=1$ supergravity.

\end{itemize}

As a caveat, our bound on the turning rate can be relaxed in the case of very large curvatures of the field space (see also \cite{Aragam:2021scu}). However, we have argued this is not a generic situation in string effective models and considered instead this contribution of order one. For example, the parameter $n$ of the metric of the hyperbolic spaces is typically very constrained by Calabi-Yau compactifications and takes very specific $\mathcal{O}(1)$ values.
Furthermore, our results should not be regarded as strictly valid only {\it at} the exact boundary of moduli space or {\it slightly away} from it (as discussed in sec.~\ref{sec::nonconstOmega}). They can be applied more generally whenever the geometry of the moduli space can be approximated as hyperbolic and to any scenario of cosmic acceleration characterized by \textit{large field excursions}, as long as the SDC is satisfied.

Finally, we acknowledge that our investigation has focused on a minimal setup and has not taken into account other details that can arise in realistic cosmological descriptions of inflation or dark energy.  One important aspect that we have not considered is the presence of other sources of energy density. One such example is provided by scalar-gauge field interactions, which lead to additional friction terms in the equations of motion. In the context of inflation, this has been studied in the seminal paper \cite{Anber:2009ua}. In the context of (quintessence) dark energy, it has been recently investigated in \cite{DallAgata:2019yrr}. Furthermore, we have studied just the homogeneous case with the scalar fields depending just on the time variable. The case of inhomogeneous  fields, i.e. $\bm{\Phi} = \bm{\Phi}\left( t, \bm{x}\right)$, can be understood as introducing extra forces in moduli space, thus again leading to deviation from geodesic trajectories. See \cite{Buratti:2018xjt} for one specific study of spatial-dependent fields and its relation to the SDC. We leave these and other exciting directions for future work.

\vspace{0.5cm}
\noindent{\bf Acknowledgments.} 
We thank J.~Calderon-Infante, M.~Cicoli, N.~Cribiori, J.~Masias, F.~G.~Pedro and I.~Zavala for useful discussions.
The work of D.L.~is supported by the Origins Excellence Cluster and by the German-Israel-Project (DIP) on Holography and the Swampland. The work of G.N. is supported by
the China Scholarship Council.

\appendix
\section{Non-affine geodesic equation}\label{appA}
Here we show that a non-affine geodesic equation can always be brought to the form of an affine geodesic equation.

This reparametrization is unique (for a given non-affine geodesic equation) up to an affine transformation (which is just a linear reparametrization, i.e. $t\to mt+n$).

Let us start with a non-affine geodesic equation. This equation has an additional term which is proportional to the first derivative, namely
\begin{equation}
	\frac{d^2x^a}{dt^2} + \Gamma_{bc}^a \frac{dx^b}{dt}\frac{dx^c}{dt} = \alpha \frac{dx^a}{dt}\,. \label{nonaffinegeod}
\end{equation}
Such a first order derivative term is usually referred to as a friction term (depending on the sign). Now we can introduce a new parameter $s(t)$ so that
\begin{equation}
	\frac{d}{dt} = \frac{ds}{dt} \frac{d}{ds}\,.
\end{equation}
Then we get
\begin{eqnarray}
	\frac{d}{dt} \frac{dx^a}{dt} &=& \frac{d}{dt} \left( \frac{ds}{dt} \frac{dx^a}{ds} \right) \\
	&=& \left( \frac{ds}{dt} \right)^2 	\frac{d^2x^a}{ds^2} + \frac{dx^a}{ds} \frac{d}{dt} \left( \frac{ds}{dt} \right)\,.
\end{eqnarray}
This allows us the rewrite the geodesic equation \eqref{nonaffinegeod} as
\begin{equation}
	 \left( \frac{ds}{dt} \right)^2 \frac{d^2x^a}{ds^2} +\left( \frac{ds}{dt} \right)^2 \Gamma_{bc}^a \frac{dx^b}{ds}\frac{dx^c}{ds} = \frac{dx^a}{ds} \left( \alpha\frac{ds}{dt}- \frac{d}{dt} \left( \frac{ds}{dt} \right) \right)\,.
\end{equation}
Therefore, if we want the right-hand side to vanish, we need to solve the following equation (setting $\lambda \equiv ds/dt$)
\begin{equation}
	\alpha \lambda = \frac{d\lambda}{dt}\,,
\end{equation}
which can be integrated and one gets
\begin{equation}
	\lambda = \lambda_0 e^{\int\!\alpha\; dt}\,.
\end{equation}
Hence, the geodesic equation \eqref{nonaffinegeod} becomes affine in the $s$-parametrization, that is
\begin{equation}
	\frac{d^2x^a}{ds^2} + \Gamma_{bc}^a \frac{dx^b}{ds}\frac{dx^c}{ds} = 0\,,
\end{equation}
which is the standard geodesic equation form and the equation defining the parallel transport of a tangent vector along itself.

We can now apply this strategy to the setup defined by eq.~\eqref{eq::eomInf}, where we have
\begin{equation}
	\frac{d^2\Phi^a}{dt^2} + \Gamma_{bc}^a \frac{d\Phi^b}{dt}\frac{d\Phi^c}{dt}+ 3H \frac{d\Phi^a}{dt} + G^{ab} V_b = 0\,.
\end{equation}
We now introduce the parameter $s(t)$. The potential term stays unchanged since it does not involve any derivative with respect to the parameter along the curve. The above equation then becomes
\begin{equation}
		\frac{d^2\Phi^a}{ds^2} + \Gamma_{bc}^a \frac{d\Phi^b}{ds}\frac{d\Phi^c}{ds}+ g^{ab} V_b = 0\,,
\end{equation}
provided we have
\begin{equation}
	\frac{ds}{dt} = \lambda = \lambda_0 e^{3Ht}\,,
\end{equation}
with $H$ approximately constant. So, we have seen that a simple reparametrization can eliminate the friction term of the equation of motion. This will have an impact on the velocity along the curve, which depends on the specific parameter.

\section{Geodesics of hyperbolic planes}
\label{appB}
 This appendix provides a complementary perspective to the results obtained in sec.~\ref{sec::Omega}. We show here that, if $\Omega = 0$ for a trajectory, then this trajectory fulfills the standard geodesic equation. We begin by examining the single hyperbolic plane which we recall here again
\begin{equation}
    \rmd\Delta^2 = \frac{n^2}{s^2} \left( \rmd s^2 + \rmd\phi^2\right)\,.
\end{equation}
This parameterizes the upper half of the hyperbolic plane, i.e. $s>0$. The non-vanishing Christoffel symbols are
\begin{equation}
    \Gamma_{ss}^s =- \frac{1}{s} =\Gamma_{s\phi}^\phi, ~~ \Gamma_{\phi\phi}^s
    =\frac{1}{s},
\end{equation}
such that the geodesic equation \eqref{eq::geodeq} or \eqref{eq::geodeqCov} becomes
\begin{eqnarray}
    \ddot{s} - \frac{1}{s} \dot{s}^2 + \frac{1}{s} \dot{\phi}^2 &=& 0\,, \\
    \ddot{\phi} - \frac{2}{s} \dot{s} \dot{\phi} &=& 0\,.
\end{eqnarray}
There are two types of geodesics which solve these equations: semi-circles with centers in $s=0$ and vertical lines with $\phi=\phi_0=\text{const}$. We are interested in the region of large $s$, so we focus on the vertical lines. For this case, the second geodesic equation becomes trivial (0=0) whereas the first reads
\begin{equation}
    \ddot{s} - \frac{1}{s} \dot{s}^2 = 0\,.
\end{equation}
After some simple algebra,  we get the solution 
\begin{eqnarray}
    s(t) &=& C ~ e^{at}\,,
\end{eqnarray}
 with $a$ and $C$ some integration constants.
We now explore what happens when we assume a constant deviation angle from a geodesic, namely, a constant velocity ratio, such as $\beta=\dot{\phi}/\dot{s}=\text{const}$. We can immediately understand that this cannot be a geodesic unless $\beta =0$, because we just learned that a (infinite-distance) geodesic must have $\dot{\phi} =0$. In this case, the set of two geodesic equations becomes
\begin{eqnarray}
    \ddot{s} - \frac{1}{s} \dot{s}^2 (1-\beta^2) &=& 0\,, \\
    \beta \left( \ddot{s} - \frac{2}{s} \dot{s}^2 \right) &=& 0\,.
\end{eqnarray}
It turns out this can only be consistently solved only for $\beta =0$. That explains what happens at the level of the geodesic equation and is in perfect agreement with the result for the turning rate \eqref{eq::OneHypOmegaInBeta}. This is also consistent with the findings in \cite{Calderon-Infante:2020dhm}.

Next, we turn to the product of two hyperbolic planes. Here the metric reads
\begin{equation}
    \rmd\Delta^2 = \frac{n^2}{s^2} \left( \rmd s^2 + \rmd\phi^2\right) +\frac{m^2}{u^2} \left( \rmd u^2 + \rmd\psi^2\right)\,.
\end{equation}
The non-vanishing Christoffel symbols are
\begin{equation}
    \Gamma_{ss}^s =- \frac{1}{s} =\Gamma_{s\phi}^\phi\,, \qquad \Gamma_{\phi\phi}^s
    =\frac{1}{s}\,, \qquad \Gamma_{uu}^u =- \frac{1}{u} =\Gamma_{s\psi}^\psi\,, \qquad  \Gamma_{\psi\psi}^u
    =\frac{1}{u}\,.
\end{equation}
Therefore, we arrive at the following geodesic equation
\begin{eqnarray}
    \ddot{s} - \frac{1}{s} \dot{s}^2 + \frac{1}{s} \dot{\phi}^2 &=& 0\,,\\
    \ddot{\phi} - \frac{2}{s} \dot{s} \dot{\phi} &=& 0\,,\\
    \ddot{u} - \frac{1}{u} \dot{u}^2 + \frac{1}{u} \dot{\psi}^2 &=& 0\,, \\
    \ddot{\psi} - \frac{2}{u} \dot{u} \dot{\psi} &=& 0\,.
\end{eqnarray}
It is pretty evident from these equations that a geodesic for the product of two hyperbolic planes consists of two geodesics of the single hyperbolic plane combined in one vector. Due to the same reasons as above we reject all the semi-circle solutions and just focus on the case with constant axions, namely $\phi=\phi_0=\text{const}$ and $\psi=\psi_0=\text{const}$. Hence, the axion equations become again trivial and we get two copies the same saxion equation which is precisely the same as above.
\begin{eqnarray}
    \ddot{s} - \frac{1}{s} \dot{s}^2  &=& 0\,,\\
    \ddot{u} - \frac{1}{u} \dot{u}^2  &=& 0\,.
\end{eqnarray}
The solutions to these equations also works out to be
\begin{eqnarray}
    s(t) &=& C e^{at}\,,\\ \label{eq::solutionaxions2Hyp}
    u(t) &=& D e^{bt}\,,
\end{eqnarray}
where $a$, $b$, $C$, $D$ are positive numbers. Now it is interesting to see what happens in the case of constant deviation between the two geodesic trajectories. This correspond to a constant ratio of velocities, namely, $\dot{u}/u = \delta\dot{s}/s$ for some $\delta = \text{const}$. We then get
\begin{eqnarray}
    \ddot{s} - \frac{1}{s} \dot{s}^2  &=& 0\,, \\
  \delta \left( \ddot{s} - \frac{1}{s} \dot{s}^2 \right)  &=& 0\,.
\end{eqnarray}
Unlike above, these two equations are of course compatible. The solution of this equation is precisely as given above in eq.~\eqref{eq::solutionaxions2Hyp}.
Now we can integrate the condition on the velocities 
\begin{equation}
    \ln u = \delta \ln s + k
\end{equation}
where k is an integration constant. Plugging the solution for $s$ into that, we arrive at
\begin{equation}
    \ln u = \delta \ln s + k = (\delta a) ~ t + (\delta C+ k)
\end{equation}
which is obviously also a solution to the geodesic equation for constant axions. Therefore, in a product of two hyperbolic spaces, any linear combination of saxionic trajectories is a geodesic, which is in agreement with the result $\Omega =0$ of sec.~\ref{sec:saxionsaxion2Hyp}.

Let us now explore what happens in the case of trajectories involving displacements of the saxion and the axion of two different hyperbolic planes, as discussed in sec.~\ref{sec:axionsaxion2Hyp}. So we consider $\phi = \phi_0 = \text{const}$ and $u = u_0 = \text{const}$. This makes the equation associated with $\phi$ trivial again. However, the equation for the other constant coordinate $u$ is not trivial and this makes a crucial difference with the previous case. Omitting the equation for the field $\phi$, we get
\begin{eqnarray}
    \ddot{s} - \frac{1}{s} \dot{s}^2 &=& 0 \,,\\
     \frac{1}{u_0} \dot{\psi}^2 &=& 0\,, \\
    \ddot{\psi}  &=& 0\,.
\end{eqnarray}
We can already read off some implications from this set of equations, namely $\psi$ has to be constant. But let us employ condition for which the trajectory is characterized by a constant ratio of the velocities, namely $\dot{\psi} = \gamma\dot{s}/{s}$. So the above equations turn into
\begin{eqnarray}
    \ddot{s} - \frac{1}{s} \dot{s}^2 &=& 0 \\
     \frac{1}{u_0} \gamma^2 \left( \frac{\dot{s}}{s} \right)^2 &=& 0 \\
    \gamma \left( \ddot{s} - \frac{1}{s} \dot{s}^2 \right)  &=& 0.
\end{eqnarray}
Without the second equation we would be in the same situation as in the previous case, which had implied $\Omega = 0$. However, precisely this second equation spoils the situation since it forces upon us the uninteresting case $s=\text{const}$. Instead, we have required our trajectory to approach $s = \infty$. So there is no solution to the geodesic equation in this case. In agreement with the findings of sec.~\ref{sec::Omega}, we conclude that $\Omega \neq 0$ and that we have to include all the coordinates in order to get a proper result.

\newpage
\bibliographystyle{JHEP}
\bibliography{SDCturnV2.bib}

\providecommand{\href}[2]{#2}\begingroup\raggedright\begin{thebibliography}{10}

\bibitem{SupernovaSearchTeam:1998fmf}
{\scshape Supernova Search Team} collaboration, A.~G. Riess et~al.,
  \emph{{Observational evidence from supernovae for an accelerating universe
  and a cosmological constant}},
  \href{http://dx.doi.org/10.1086/300499}{\emph{Astron. J.} {\bf 116} (1998)
  1009--1038}, [\href{https://arxiv.org/abs/astro-ph/9805201}{{\tt
  astro-ph/9805201}}].

\bibitem{Riess:2021jrx}
A.~G. Riess et~al., \emph{{A Comprehensive Measurement of the Local Value of
  the Hubble Constant with 1 km/s/Mpc Uncertainty from the Hubble Space
  Telescope and the SH0ES Team}},  \href{https://arxiv.org/abs/2112.04510}{{\tt
  2112.04510}}.

\bibitem{Boomerang:2000jdg}
{\scshape Boomerang} collaboration, A.~H. Jaffe et~al., \emph{{Cosmology from
  MAXIMA-1, BOOMERANG and COBE / DMR CMB observations}},
  \href{http://dx.doi.org/10.1103/PhysRevLett.86.3475}{\emph{Phys. Rev. Lett.}
  {\bf 86} (2001) 3475--3479},
  [\href{https://arxiv.org/abs/astro-ph/0007333}{{\tt astro-ph/0007333}}].

\bibitem{SDSS:2003eyi}
{\scshape SDSS} collaboration, M.~Tegmark et~al., \emph{{Cosmological
  parameters from SDSS and WMAP}},
  \href{http://dx.doi.org/10.1103/PhysRevD.69.103501}{\emph{Phys. Rev. D} {\bf
  69} (2004) 103501}, [\href{https://arxiv.org/abs/astro-ph/0310723}{{\tt
  astro-ph/0310723}}].

\bibitem{Aghanim:2018eyx}
{\scshape Planck} collaboration, N.~Aghanim et~al., \emph{{Planck 2018 results.
  VI. Cosmological parameters}},  \href{https://arxiv.org/abs/1807.06209}{{\tt
  1807.06209}}.

\bibitem{Cicoli:2023opf}
M.~Cicoli, J.~P. Conlon, A.~Maharana, S.~Parameswaran, F.~Quevedo and
  I.~Zavala, \emph{{String Cosmology: from the Early Universe to Today}},
  \href{https://arxiv.org/abs/2303.04819}{{\tt 2303.04819}}.

\bibitem{Chakraborty:2019dfh}
D.~Chakraborty, R.~Chiovoloni, O.~Loaiza-Brito, G.~Niz and I.~Zavala,
  \emph{{Fat inflatons, large turns and the $\eta$-problem}},
  \href{http://dx.doi.org/10.1088/1475-7516/2020/01/020}{\emph{JCAP} {\bf 01}
  (2020) 020}, [\href{https://arxiv.org/abs/1908.09797}{{\tt 1908.09797}}].

\bibitem{Dimopoulos:2005ac}
S.~Dimopoulos, S.~Kachru, J.~McGreevy and J.~G. Wacker, \emph{{N-flation}},
  \href{http://dx.doi.org/10.1088/1475-7516/2008/08/003}{\emph{JCAP} {\bf 08}
  (2008) 003}, [\href{https://arxiv.org/abs/hep-th/0507205}{{\tt
  hep-th/0507205}}].

\bibitem{Wands:2007bd}
D.~Wands, \emph{{Multiple field inflation}},
  \href{http://dx.doi.org/10.1007/978-3-540-74353-8_8}{\emph{Lect. Notes Phys.}
  {\bf 738} (2008) 275--304},
  [\href{https://arxiv.org/abs/astro-ph/0702187}{{\tt astro-ph/0702187}}].

\bibitem{Cremonini:2010ua}
S.~Cremonini, Z.~Lalak and K.~Turzynski, \emph{{Strongly Coupled Perturbations
  in Two-Field Inflationary Models}},
  \href{http://dx.doi.org/10.1088/1475-7516/2011/03/016}{\emph{JCAP} {\bf 03}
  (2011) 016}, [\href{https://arxiv.org/abs/1010.3021}{{\tt 1010.3021}}].

\bibitem{Yang:2012bs}
I.-S. Yang, \emph{{The Strong Multifield Slowroll Condition and Spiral
  Inflation}}, \href{http://dx.doi.org/10.1103/PhysRevD.85.123532}{\emph{Phys.
  Rev. D} {\bf 85} (2012) 123532}, [\href{https://arxiv.org/abs/1202.3388}{{\tt
  1202.3388}}].

\bibitem{Brown:2017osf}
A.~R. Brown, \emph{{Hyperbolic Inflation}},
  \href{http://dx.doi.org/10.1103/PhysRevLett.121.251601}{\emph{Phys. Rev.
  Lett.} {\bf 121} (2018) 251601},
  [\href{https://arxiv.org/abs/1705.03023}{{\tt 1705.03023}}].

\bibitem{Christodoulidis:2018qdw}
P.~Christodoulidis, D.~Roest and E.~I. Sfakianakis, \emph{{Angular inflation in
  multi-field $\alpha$-attractors}},
  \href{http://dx.doi.org/10.1088/1475-7516/2019/11/002}{\emph{JCAP} {\bf 11}
  (2019) 002}, [\href{https://arxiv.org/abs/1803.09841}{{\tt 1803.09841}}].

\bibitem{Dias:2018pgj}
M.~Dias, J.~Frazer, A.~Retolaza, M.~Scalisi and A.~Westphal, \emph{{Pole
  N-flation}}, \href{http://dx.doi.org/10.1007/JHEP02(2019)120}{\emph{JHEP}
  {\bf 02} (2019) 120}, [\href{https://arxiv.org/abs/1805.02659}{{\tt
  1805.02659}}].

\bibitem{Achucarro:2018vey}
A.~Ach\'ucarro and G.~A. Palma, \emph{{The string swampland constraints require
  multi-field inflation}},
  \href{http://dx.doi.org/10.1088/1475-7516/2019/02/041}{\emph{JCAP} {\bf 02}
  (2019) 041}, [\href{https://arxiv.org/abs/1807.04390}{{\tt 1807.04390}}].

\bibitem{Aragam:2019omo}
V.~Aragam, S.~Paban and R.~Rosati, \emph{{Multi-field Inflation in High-Slope
  Potentials}},
  \href{http://dx.doi.org/10.1088/1475-7516/2020/04/022}{\emph{JCAP} {\bf 04}
  (2020) 022}, [\href{https://arxiv.org/abs/1905.07495}{{\tt 1905.07495}}].

\bibitem{Aragam:2020uqi}
V.~Aragam, S.~Paban and R.~Rosati, \emph{{The Multi-Field, Rapid-Turn
  Inflationary Solution}},
  \href{http://dx.doi.org/10.1007/JHEP03(2021)009}{\emph{JHEP} {\bf 03} (2021)
  009}, [\href{https://arxiv.org/abs/2010.15933}{{\tt 2010.15933}}].

\bibitem{Aragam:2021scu}
V.~Aragam, R.~Chiovoloni, S.~Paban, R.~Rosati and I.~Zavala, \emph{{Rapid-turn
  inflation in supergravity is rare and tachyonic}},
  \href{http://dx.doi.org/10.1088/1475-7516/2022/03/002}{\emph{JCAP} {\bf 03}
  (2022) 002}, [\href{https://arxiv.org/abs/2110.05516}{{\tt 2110.05516}}].

\bibitem{Renaux-Petel:2021yxh}
S.~Renaux-Petel, \emph{{Inflation with strongly non-geodesic motion:
  theoretical motivations and observational imprints}},
  \href{http://dx.doi.org/10.22323/1.398.0128}{\emph{PoS} {\bf EPS-HEP2021}
  (2022) 128}, [\href{https://arxiv.org/abs/2111.00989}{{\tt 2111.00989}}].

\bibitem{Bhattacharya:2022fze}
S.~Bhattacharya and I.~Zavala, \emph{{Sharp turns in axion monodromy:
  primordial black holes and gravitational waves}},
  \href{http://dx.doi.org/10.1088/1475-7516/2023/04/065}{\emph{JCAP} {\bf 04}
  (2023) 065}, [\href{https://arxiv.org/abs/2205.06065}{{\tt 2205.06065}}].

\bibitem{Cicoli:2020cfj}
M.~Cicoli, G.~Dibitetto and F.~G. Pedro, \emph{{New accelerating solutions in
  late-time cosmology}},
  \href{http://dx.doi.org/10.1103/PhysRevD.101.103524}{\emph{Phys. Rev. D} {\bf
  101} (2020) 103524}, [\href{https://arxiv.org/abs/2002.02695}{{\tt
  2002.02695}}].

\bibitem{Cicoli:2020noz}
M.~Cicoli, G.~Dibitetto and F.~G. Pedro, \emph{{Out of the Swampland with
  Multifield Quintessence?}},
  \href{http://dx.doi.org/10.1007/JHEP10(2020)035}{\emph{JHEP} {\bf 10} (2020)
  035}, [\href{https://arxiv.org/abs/2007.11011}{{\tt 2007.11011}}].

\bibitem{Akrami:2020zfz}
Y.~Akrami, M.~Sasaki, A.~R. Solomon and V.~Vardanyan, \emph{{Multi-field dark
  energy: Cosmic acceleration on a steep potential}},
  \href{http://dx.doi.org/10.1016/j.physletb.2021.136427}{\emph{Phys. Lett. B}
  {\bf 819} (2021) 136427}, [\href{https://arxiv.org/abs/2008.13660}{{\tt
  2008.13660}}].

\bibitem{Anguelova:2021jxu}
L.~Anguelova, J.~Dumancic, R.~Gass and L.~C.~R. Wijewardhana, \emph{{Dark
  energy from inspiraling in field space}},
  \href{http://dx.doi.org/10.1088/1475-7516/2022/03/018}{\emph{JCAP} {\bf 03}
  (2022) 018}, [\href{https://arxiv.org/abs/2111.12136}{{\tt 2111.12136}}].

\bibitem{Eskilt:2022zky}
J.~R. Eskilt, Y.~Akrami, A.~R. Solomon and V.~Vardanyan, \emph{{Cosmological
  dynamics of multifield dark energy}},
  \href{http://dx.doi.org/10.1103/PhysRevD.106.023512}{\emph{Phys. Rev. D} {\bf
  106} (2022) 023512}, [\href{https://arxiv.org/abs/2201.08841}{{\tt
  2201.08841}}].

\bibitem{Brinkmann:2022oxy}
M.~Brinkmann, M.~Cicoli, G.~Dibitetto and F.~G. Pedro, \emph{{Stringy
  multifield quintessence and the Swampland}},
  \href{http://dx.doi.org/10.1007/JHEP11(2022)044}{\emph{JHEP} {\bf 11} (2022)
  044}, [\href{https://arxiv.org/abs/2206.10649}{{\tt 2206.10649}}].

\bibitem{Shiu:2023nph}
G.~Shiu, F.~Tonioni and H.~V. Tran, \emph{{Accelerating universe at the end of
  time}},  \href{https://arxiv.org/abs/2303.03418}{{\tt 2303.03418}}.

\bibitem{Shiu:2023rxt}
G.~Shiu, F.~Tonioni and H.~V. Tran, \emph{{Late-time attractors and cosmic
  acceleration}},  \href{https://arxiv.org/abs/2306.07327}{{\tt 2306.07327}}.

\bibitem{Palma:2020ejf}
G.~A. Palma, S.~Sypsas and C.~Zenteno, \emph{{Seeding primordial black holes in
  multifield inflation}},
  \href{http://dx.doi.org/10.1103/PhysRevLett.125.121301}{\emph{Phys. Rev.
  Lett.} {\bf 125} (2020) 121301},
  [\href{https://arxiv.org/abs/2004.06106}{{\tt 2004.06106}}].

\bibitem{Fumagalli:2020adf}
J.~Fumagalli, S.~Renaux-Petel, J.~W. Ronayne and L.~T. Witkowski,
  \emph{{Turning in the landscape: A new mechanism for generating primordial
  black holes}},
  \href{http://dx.doi.org/10.1016/j.physletb.2023.137921}{\emph{Phys. Lett. B}
  {\bf 841} (2023) 137921}, [\href{https://arxiv.org/abs/2004.08369}{{\tt
  2004.08369}}].

\bibitem{Anguelova:2020nzl}
L.~Anguelova, \emph{{On Primordial Black Holes from Rapid Turns in Two-field
  Models}}, \href{http://dx.doi.org/10.1088/1475-7516/2021/06/004}{\emph{JCAP}
  {\bf 06} (2021) 004}, [\href{https://arxiv.org/abs/2012.03705}{{\tt
  2012.03705}}].

\bibitem{Vafa:2005ui}
C.~Vafa, \emph{{The String landscape and the swampland}},
  \href{https://arxiv.org/abs/hep-th/0509212}{{\tt hep-th/0509212}}.

\bibitem{Ooguri:2006in}
H.~Ooguri and C.~Vafa, \emph{{On the Geometry of the String Landscape and the
  Swampland}},
  \href{http://dx.doi.org/10.1016/j.nuclphysb.2006.10.033}{\emph{Nucl. Phys. B}
  {\bf 766} (2007) 21--33}, [\href{https://arxiv.org/abs/hep-th/0605264}{{\tt
  hep-th/0605264}}].

\bibitem{Palti:2019pca}
E.~Palti, \emph{{The Swampland: Introduction and Review}},
  \href{http://dx.doi.org/10.1002/prop.201900037}{\emph{Fortsch. Phys.} {\bf
  67} (2019) 1900037}, [\href{https://arxiv.org/abs/1903.06239}{{\tt
  1903.06239}}].

\bibitem{vanBeest:2021lhn}
M.~van Beest, J.~Calder\'on-Infante, D.~Mirfendereski and I.~Valenzuela,
  \emph{{Lectures on the Swampland Program in String Compactifications}},
  \href{http://dx.doi.org/10.1016/j.physrep.2022.09.002}{\emph{Phys. Rept.}
  {\bf 989} (2022) 1--50}, [\href{https://arxiv.org/abs/2102.01111}{{\tt
  2102.01111}}].

\bibitem{Dvali:2007hz}
G.~Dvali, \emph{{Black Holes and Large N Species Solution to the Hierarchy
  Problem}}, \href{http://dx.doi.org/10.1002/prop.201000009}{\emph{Fortsch.
  Phys.} {\bf 58} (2010) 528--536},
  [\href{https://arxiv.org/abs/0706.2050}{{\tt 0706.2050}}].

\bibitem{Dvali:2007wp}
G.~Dvali and M.~Redi, \emph{{Black Hole Bound on the Number of Species and
  Quantum Gravity at LHC}},
  \href{http://dx.doi.org/10.1103/PhysRevD.77.045027}{\emph{Phys. Rev. D} {\bf
  77} (2008) 045027}, [\href{https://arxiv.org/abs/0710.4344}{{\tt
  0710.4344}}].

\bibitem{Dvali:2009ks}
G.~Dvali and D.~{L\"ust}, \emph{{Evaporation of Microscopic Black Holes in
  String Theory and the Bound on Species}},
  \href{http://dx.doi.org/10.1002/prop.201000008}{\emph{Fortsch. Phys.} {\bf
  58} (2010) 505--527}, [\href{https://arxiv.org/abs/0912.3167}{{\tt
  0912.3167}}].

\bibitem{Dvali:2010vm}
G.~Dvali and C.~Gomez, \emph{{Species and Strings}},
  \href{https://arxiv.org/abs/1004.3744}{{\tt 1004.3744}}.

\bibitem{Dvali:2012uq}
G.~Dvali, C.~Gomez and D.~{L\"ust}, \emph{{Black Hole Quantum Mechanics in the
  Presence of Species}},
  \href{http://dx.doi.org/10.1002/prop.201300002}{\emph{Fortsch. Phys.} {\bf
  61} (2013) 768--778}, [\href{https://arxiv.org/abs/1206.2365}{{\tt
  1206.2365}}].

\bibitem{Lust:2019zwm}
D.~L\"ust, E.~Palti and C.~Vafa, \emph{{AdS and the Swampland}},
  \href{http://dx.doi.org/10.1016/j.physletb.2019.134867}{\emph{Phys. Lett. B}
  {\bf 797} (2019) 134867}, [\href{https://arxiv.org/abs/1906.05225}{{\tt
  1906.05225}}].

\bibitem{Cribiori:2021gbf}
N.~Cribiori, D.~{L\"ust} and M.~Scalisi, \emph{{The gravitino and the
  swampland}}, \href{http://dx.doi.org/10.1007/JHEP06(2021)071}{\emph{JHEP}
  {\bf 06} (2021) 071}, [\href{https://arxiv.org/abs/2104.08288}{{\tt
  2104.08288}}].

\bibitem{Castellano:2021yye}
A.~Castellano, A.~Font, A.~Herraez and L.~E. Ib\'a\~nez, \emph{{A gravitino
  distance conjecture}},
  \href{http://dx.doi.org/10.1007/JHEP08(2021)092}{\emph{JHEP} {\bf 08} (2021)
  092}, [\href{https://arxiv.org/abs/2104.10181}{{\tt 2104.10181}}].

\bibitem{Anchordoqui:2023oqm}
L.~A. Anchordoqui, I.~Antoniadis, N.~Cribiori, D.~{L\"ust} and M.~Scalisi,
  \emph{{The Scale of Supersymmetry Breaking and the Dark Dimension}},
  \href{http://dx.doi.org/10.1007/JHEP05(2023)060}{\emph{JHEP} {\bf 05} (2023)
  060}, [\href{https://arxiv.org/abs/2301.07719}{{\tt 2301.07719}}].

\bibitem{Bonnefoy:2019nzv}
Q.~Bonnefoy, L.~Ciambelli, D.~L\"ust and S.~L\"ust, \emph{{Infinite Black Hole
  Entropies at Infinite Distances and Tower of States}},
  \href{http://dx.doi.org/10.1016/j.nuclphysb.2020.115112}{\emph{Nucl. Phys. B}
  {\bf 958} (2020) 115112}, [\href{https://arxiv.org/abs/1912.07453}{{\tt
  1912.07453}}].

\bibitem{Cribiori:2022cho}
N.~Cribiori, M.~Dierigl, A.~Gnecchi, D.~{L\"ust} and M.~Scalisi, \emph{{Large
  and small non-extremal black holes, thermodynamic dualities, and the
  Swampland}}, \href{http://dx.doi.org/10.1007/JHEP10(2022)093}{\emph{JHEP}
  {\bf 10} (2022) 093}, [\href{https://arxiv.org/abs/2202.04657}{{\tt
  2202.04657}}].

\bibitem{Delgado:2022dkz}
M.~Delgado, M.~Montero and C.~Vafa, \emph{{Black holes as probes of moduli
  space geometry}},
  \href{http://dx.doi.org/10.1007/JHEP04(2023)045}{\emph{JHEP} {\bf 04} (2023)
  045}, [\href{https://arxiv.org/abs/2212.08676}{{\tt 2212.08676}}].

\bibitem{Cribiori:2023ffn}
N.~Cribiori, D.~{L\"ust} and C.~Montella, \emph{{Species Entropy and
  Thermodynamics}},  \href{https://arxiv.org/abs/2305.10489}{{\tt 2305.10489}}.

\bibitem{Baume:2016psm}
F.~Baume and E.~Palti, \emph{{Backreacted Axion Field Ranges in String
  Theory}}, \href{http://dx.doi.org/10.1007/JHEP08(2016)043}{\emph{JHEP} {\bf
  08} (2016) 043}, [\href{https://arxiv.org/abs/1602.06517}{{\tt 1602.06517}}].

\bibitem{Klaewer:2016kiy}
D.~Klaewer and E.~Palti, \emph{{Super-Planckian Spatial Field Variations and
  Quantum Gravity}},
  \href{http://dx.doi.org/10.1007/JHEP01(2017)088}{\emph{JHEP} {\bf 01} (2017)
  088}, [\href{https://arxiv.org/abs/1610.00010}{{\tt 1610.00010}}].

\bibitem{Rudelius:2023mjy}
T.~Rudelius, \emph{{Revisiting the Refined Distance Conjecture}},
  \href{https://arxiv.org/abs/2303.12103}{{\tt 2303.12103}}.

\bibitem{Scalisi:2018eaz}
M.~Scalisi and I.~Valenzuela, \emph{{Swampland distance conjecture, inflation
  and $\alpha$-attractors}},
  \href{http://dx.doi.org/10.1007/JHEP08(2019)160}{\emph{JHEP} {\bf 08} (2019)
  160}, [\href{https://arxiv.org/abs/1812.07558}{{\tt 1812.07558}}].

\bibitem{Scalisi:2019gfv}
M.~Scalisi, \emph{{Inflation, Higher Spins and the Swampland}},
  \href{http://dx.doi.org/10.1016/j.physletb.2020.135683}{\emph{Phys. Lett. B}
  {\bf 808} (2020) 135683}, [\href{https://arxiv.org/abs/1912.04283}{{\tt
  1912.04283}}].

\bibitem{Bravo:2020wdr}
R.~Bravo, G.~A. Palma and S.~Riquelme, \emph{{A Tip for Landscape Riders:
  Multi-Field Inflation Can Fulfill the Swampland Distance Conjecture}},
  \href{http://dx.doi.org/10.1088/1475-7516/2020/02/004}{\emph{JCAP} {\bf 02}
  (2020) 004}, [\href{https://arxiv.org/abs/1906.05772}{{\tt 1906.05772}}].

\bibitem{Etheredge:2022opl}
M.~Etheredge, B.~Heidenreich, S.~Kaya, Y.~Qiu and T.~Rudelius,
  \emph{{Sharpening the Distance Conjecture in diverse dimensions}},
  \href{http://dx.doi.org/10.1007/JHEP12(2022)114}{\emph{JHEP} {\bf 12} (2022)
  114}, [\href{https://arxiv.org/abs/2206.04063}{{\tt 2206.04063}}].

\bibitem{vandeHeisteeg:2023uxj}
D.~van~de Heisteeg, C.~Vafa, M.~Wiesner and D.~H. Wu, \emph{{Bounds on Field
  Range for Slowly Varying Positive Potentials}},
  \href{https://arxiv.org/abs/2305.07701}{{\tt 2305.07701}}.

\bibitem{Roest:2013aoa}
D.~Roest, M.~Scalisi and I.~Zavala, \emph{{K\"ahler potentials for Planck
  inflation}},
  \href{http://dx.doi.org/10.1088/1475-7516/2013/11/007}{\emph{JCAP} {\bf 11}
  (2013) 007}, [\href{https://arxiv.org/abs/1307.4343}{{\tt 1307.4343}}].

\bibitem{Kallosh:2013yoa}
R.~Kallosh, A.~Linde and D.~Roest, \emph{{Superconformal Inflationary
  $\alpha$-Attractors}},
  \href{http://dx.doi.org/10.1007/JHEP11(2013)198}{\emph{JHEP} {\bf 11} (2013)
  198}, [\href{https://arxiv.org/abs/1311.0472}{{\tt 1311.0472}}].

\bibitem{Burgess:2014tja}
C.~P. Burgess, M.~Cicoli, F.~Quevedo and M.~Williams, \emph{{Inflating with
  Large Effective Fields}},
  \href{http://dx.doi.org/10.1088/1475-7516/2014/11/045}{\emph{JCAP} {\bf 11}
  (2014) 045}, [\href{https://arxiv.org/abs/1404.6236}{{\tt 1404.6236}}].

\bibitem{Burgess:2014oma}
C.~Burgess and D.~Roest, \emph{{Inflation by Alignment}},
  \href{http://dx.doi.org/10.1088/1475-7516/2015/06/012}{\emph{JCAP} {\bf 06}
  (2015) 012}, [\href{https://arxiv.org/abs/1412.1614}{{\tt 1412.1614}}].

\bibitem{Roest:2015qya}
D.~Roest and M.~Scalisi, \emph{{Cosmological attractors from
  \ensuremath{\alpha}-scale supergravity}},
  \href{http://dx.doi.org/10.1103/PhysRevD.92.043525}{\emph{Phys. Rev. D} {\bf
  92} (2015) 043525}, [\href{https://arxiv.org/abs/1503.07909}{{\tt
  1503.07909}}].

\bibitem{Burgess:2020qsc}
C.~P. Burgess, M.~Cicoli, D.~Ciupke, S.~Krippendorf and F.~Quevedo, \emph{{UV
  Shadows in EFTs: Accidental Symmetries, Robustness and No-Scale
  Supergravity}},
  \href{http://dx.doi.org/10.1002/prop.202000076}{\emph{Fortsch. Phys.} {\bf
  68} (2020) 2000076}, [\href{https://arxiv.org/abs/2006.06694}{{\tt
  2006.06694}}].

\bibitem{Ooguri:2018wrx}
H.~Ooguri, E.~Palti, G.~Shiu and C.~Vafa, \emph{{Distance and de Sitter
  Conjectures on the Swampland}},
  \href{http://dx.doi.org/10.1016/j.physletb.2018.11.018}{\emph{Phys. Lett. B}
  {\bf 788} (2019) 180--184}, [\href{https://arxiv.org/abs/1810.05506}{{\tt
  1810.05506}}].

\bibitem{Hebecker:2018vxz}
A.~Hebecker and T.~Wrase, \emph{{The Asymptotic dS Swampland Conjecture - a
  Simplified Derivation and a Potential Loophole}},
  \href{http://dx.doi.org/10.1002/prop.201800097}{\emph{Fortsch. Phys.} {\bf
  67} (2019) 1800097}, [\href{https://arxiv.org/abs/1810.08182}{{\tt
  1810.08182}}].

\bibitem{Rudelius:2021azq}
T.~Rudelius, \emph{{Asymptotic observables and the swampland}},
  \href{http://dx.doi.org/10.1103/PhysRevD.104.126023}{\emph{Phys. Rev. D} {\bf
  104} (2021) 126023}, [\href{https://arxiv.org/abs/2106.09026}{{\tt
  2106.09026}}].

\bibitem{Rudelius:2022gbz}
T.~Rudelius, \emph{{Asymptotic scalar field cosmology in string theory}},
  \href{http://dx.doi.org/10.1007/JHEP10(2022)018}{\emph{JHEP} {\bf 10} (2022)
  018}, [\href{https://arxiv.org/abs/2208.08989}{{\tt 2208.08989}}].

\bibitem{Calderon-Infante:2022nxb}
J.~Calder\'on-Infante, I.~Ruiz and I.~Valenzuela, \emph{{Asymptotic Accelerated
  Expansion in String Theory and the Swampland}},
  \href{https://arxiv.org/abs/2209.11821}{{\tt 2209.11821}}.

\bibitem{Marconnet:2022fmx}
P.~Marconnet and D.~Tsimpis, \emph{{Universal accelerating cosmologies from 10d
  supergravity}}, \href{http://dx.doi.org/10.1007/JHEP01(2023)033}{\emph{JHEP}
  {\bf 01} (2023) 033}, [\href{https://arxiv.org/abs/2210.10813}{{\tt
  2210.10813}}].

\bibitem{Apers:2022cyl}
F.~Apers, J.~P. Conlon, M.~Mosny and F.~Revello, \emph{{Kination, Meet Kasner:
  On The Asymptotic Cosmology of String Compactifications}},
  \href{https://arxiv.org/abs/2212.10293}{{\tt 2212.10293}}.

\bibitem{Cremonini:2023suw}
S.~Cremonini, E.~Gonzalo, M.~Rajaguru, Y.~Tang and T.~Wrase, \emph{{On
  Asymptotic Dark Energy in String Theory}},
  \href{https://arxiv.org/abs/2306.15714}{{\tt 2306.15714}}.

\bibitem{Obied:2018sgi}
G.~Obied, H.~Ooguri, L.~Spodyneiko and C.~Vafa, \emph{{De Sitter Space and the
  Swampland}},  \href{https://arxiv.org/abs/1806.08362}{{\tt 1806.08362}}.

\bibitem{Garg:2018reu}
S.~K. Garg and C.~Krishnan, \emph{{Bounds on Slow Roll and the de Sitter
  Swampland}}, \href{http://dx.doi.org/10.1007/JHEP11(2019)075}{\emph{JHEP}
  {\bf 11} (2019) 075}, [\href{https://arxiv.org/abs/1807.05193}{{\tt
  1807.05193}}].

\bibitem{Andriot:2020lea}
D.~Andriot, N.~Cribiori and D.~Erkinger, \emph{{The web of swampland
  conjectures and the TCC bound}},
  \href{http://dx.doi.org/10.1007/JHEP07(2020)162}{\emph{JHEP} {\bf 07} (2020)
  162}, [\href{https://arxiv.org/abs/2004.00030}{{\tt 2004.00030}}].

\bibitem{Gendler:2020dfp}
N.~Gendler and I.~Valenzuela, \emph{{Merging the weak gravity and distance
  conjectures using BPS extremal black holes}},
  \href{http://dx.doi.org/10.1007/JHEP01(2021)176}{\emph{JHEP} {\bf 01} (2021)
  176}, [\href{https://arxiv.org/abs/2004.10768}{{\tt 2004.10768}}].

\bibitem{Lanza:2021udy}
S.~Lanza, F.~Marchesano, L.~Martucci and I.~Valenzuela, \emph{{The EFT stringy
  viewpoint on large distances}},
  \href{http://dx.doi.org/10.1007/JHEP09(2021)197}{\emph{JHEP} {\bf 09} (2021)
  197}, [\href{https://arxiv.org/abs/2104.05726}{{\tt 2104.05726}}].

\bibitem{Grimm:2018ohb}
T.~W. Grimm, E.~Palti and I.~Valenzuela, \emph{{Infinite Distances in Field
  Space and Massless Towers of States}},
  \href{http://dx.doi.org/10.1007/JHEP08(2018)143}{\emph{JHEP} {\bf 08} (2018)
  143}, [\href{https://arxiv.org/abs/1802.08264}{{\tt 1802.08264}}].

\bibitem{Blumenhagen:2015qda}
R.~Blumenhagen, A.~Font, M.~Fuchs, D.~Herschmann and E.~Plauschinn,
  \emph{{Towards Axionic Starobinsky-like Inflation in String Theory}},
  \href{http://dx.doi.org/10.1016/j.physletb.2015.05.001}{\emph{Phys. Lett. B}
  {\bf 746} (2015) 217--222}, [\href{https://arxiv.org/abs/1503.01607}{{\tt
  1503.01607}}].

\bibitem{Valenzuela:2016yny}
I.~Valenzuela, \emph{{Backreaction Issues in Axion Monodromy and Minkowski
  4-forms}}, \href{http://dx.doi.org/10.1007/JHEP06(2017)098}{\emph{JHEP} {\bf
  06} (2017) 098}, [\href{https://arxiv.org/abs/1611.00394}{{\tt 1611.00394}}].

\bibitem{Blumenhagen:2018hsh}
R.~Blumenhagen, \emph{{Large Field Inflation/Quintessence and the Refined
  Swampland Distance Conjecture}},
  \href{http://dx.doi.org/10.22323/1.318.0175}{\emph{PoS} {\bf CORFU2017}
  (2018) 175}, [\href{https://arxiv.org/abs/1804.10504}{{\tt 1804.10504}}].

\bibitem{Grimm:2020ouv}
T.~W. Grimm and C.~Li, \emph{{Universal axion backreaction in flux
  compactifications}},
  \href{http://dx.doi.org/10.1007/JHEP06(2021)067}{\emph{JHEP} {\bf 06} (2021)
  067}, [\href{https://arxiv.org/abs/2012.08272}{{\tt 2012.08272}}].

\bibitem{Calderon-Infante:2020dhm}
J.~Calder\'on-Infante, A.~M. Uranga and I.~Valenzuela, \emph{{The Convex Hull
  Swampland Distance Conjecture and Bounds on Non-geodesics}},
  \href{http://dx.doi.org/10.1007/JHEP03(2021)299}{\emph{JHEP} {\bf 03} (2021)
  299}, [\href{https://arxiv.org/abs/2012.00034}{{\tt 2012.00034}}].

\bibitem{Bjorkmo:2019aev}
T.~Bjorkmo and M.~C.~D. Marsh, \emph{{Hyperinflation generalised: from its
  attractor mechanism to its tension with the \textquoteleft{}swampland
  conditions\textquoteright{}}},
  \href{http://dx.doi.org/10.1007/JHEP04(2019)172}{\emph{JHEP} {\bf 04} (2019)
  172}, [\href{https://arxiv.org/abs/1901.08603}{{\tt 1901.08603}}].

\bibitem{Scalisi:2015qga}
M.~Scalisi, \emph{{Cosmological $\alpha$-attractors and de Sitter landscape}},
  \href{http://dx.doi.org/10.1007/JHEP12(2015)134}{\emph{JHEP} {\bf 12} (2015)
  134}, [\href{https://arxiv.org/abs/1506.01368}{{\tt 1506.01368}}].

\bibitem{Tasinato:2023ukp}
G.~Tasinato, \emph{{A large $|\eta|$ approach to single field inflation}},
  \href{https://arxiv.org/abs/2305.11568}{{\tt 2305.11568}}.

\bibitem{Lee:2019wij}
S.-J. Lee, W.~Lerche and T.~Weigand, \emph{{Emergent strings from infinite
  distance limits}},
  \href{http://dx.doi.org/10.1007/JHEP02(2022)190}{\emph{JHEP} {\bf 02} (2022)
  190}, [\href{https://arxiv.org/abs/1910.01135}{{\tt 1910.01135}}].

\bibitem{vandeHeisteeg:2023ubh}
D.~van~de Heisteeg, C.~Vafa and M.~Wiesner, \emph{{Bounds on Species Scale and
  the Distance Conjecture}},  \href{https://arxiv.org/abs/2303.13580}{{\tt
  2303.13580}}.

\bibitem{Hetz:2016ics}
A.~Hetz and G.~A. Palma, \emph{{Sound Speed of Primordial Fluctuations in
  Supergravity Inflation}},
  \href{http://dx.doi.org/10.1103/PhysRevLett.117.101301}{\emph{Phys. Rev.
  Lett.} {\bf 117} (2016) 101301},
  [\href{https://arxiv.org/abs/1601.05457}{{\tt 1601.05457}}].

\bibitem{Anber:2009ua}
M.~M. Anber and L.~Sorbo, \emph{{Naturally inflating on steep potentials
  through electromagnetic dissipation}},
  \href{http://dx.doi.org/10.1103/PhysRevD.81.043534}{\emph{Phys. Rev. D} {\bf
  81} (2010) 043534}, [\href{https://arxiv.org/abs/0908.4089}{{\tt
  0908.4089}}].

\bibitem{DallAgata:2019yrr}
G.~Dall'Agata, S.~Gonz\'alez-Mart\'\i{}n, A.~Papageorgiou and M.~Peloso,
  \emph{{Warm dark energy}},
  \href{http://dx.doi.org/10.1088/1475-7516/2020/08/032}{\emph{JCAP} {\bf 08}
  (2020) 032}, [\href{https://arxiv.org/abs/1912.09950}{{\tt 1912.09950}}].

\bibitem{Buratti:2018xjt}
G.~Buratti, J.~Calder\'on and A.~M. Uranga, \emph{{Transplanckian axion
  monodromy!?}}, \href{http://dx.doi.org/10.1007/JHEP05(2019)176}{\emph{JHEP}
  {\bf 05} (2019) 176}, [\href{https://arxiv.org/abs/1812.05016}{{\tt
  1812.05016}}].

\end{thebibliography}\endgroup

\end{document}